\documentclass{aa}
\usepackage{subcaption}
\usepackage{graphicx}
\usepackage{lastpage}
\usepackage{float}
\usepackage{xcolor}
\RequirePackage{etex}

\usepackage{txfonts}

\usepackage[
    colorlinks=true,       
    linkcolor=blue,        
    citecolor=blue,        
    urlcolor=blue          
]{hyperref}
\usepackage{booktabs}
\usepackage{xcolor}

\makeatletter

\renewcommand*\aa@manuscriptname{%
  manuscript no. aa61285-26
  \hspace{\stretch{1}}%
  \copyright ESO \the\year
}

\renewcommand*\aa@textidlineempty{%
  {\slshape A\&A proofs:}\ manuscript no.~aa61285-26
}

\makeatother
\begin{document}
\title{Magnetic flux cancellation in a 3D magnetohydrodynamic simulation of an emerging flux tube with a non-uniform twist}

\author{V. Agalianou\inst{1}\thanks{email:v.agalianou@uoi.gr}, S. Patsourakos\inst{1}, V. Archontis\inst{1}
}
\institute{Department of Physics, University of Ioannina,
Ioannina, Greece\\
}


 
  \abstract
{Flux emergence is often accompanied by flux cancellation, which may drive dynamic phenomena in the solar atmosphere. In numerical models, emerging flux tubes with a non-uniform twist can produce complex multipolar photospheric patterns, in which flux cancellation can occur.}
  {We studied magnetic flux cancellation in a three-dimensional resistive magnetohydrodynamic (MHD) simulation of an emerging flux tube with a non-uniform twist profile.} 
{We solved time-dependent resistive MHD equations to model the emergence of a magnetic flux tube in a magnetised background atmosphere. 
We identified several flux cancellation sites, estimated their cancellation rates, and focused mainly on the strongest event. We examined the topology of the magnetic fields in the low atmosphere and associated the flux cancellation event with a blowout jet that occurred during the simulation.}
{The configuration of the emerging field does not follow the simple emergence of a dipolar field. Instead, multiple opposite polarities appeared due to the non-uniform twist profile of the emerging flux tube, leading to a complex multipolar photospheric structure. We detected interactions between inner secondary magnetic patches, which resulted in flux cancellation of the photospheric $\mathrm{B_z}$ along the polarity inversion line (PIL) of the inner secondary magnetic elements. These interactions led to the formation of concave-up loops and a cancellation rate of \( 1.1 \times 10^{19}\ \mathrm{Mx\ hr^{-1}} \). During the flux cancellation process, magnetic reconnection between converging opposite polarities progressively created a coherent magnetic flux rope above the cancellation region. The evolution of the rope is associated with the blowout jet, which occurs as the rope expands and releases part of its stored twist into the overlying corona. We also briefly examined two smaller secondary cancellation events detected at photospheric heights. The results we report highlight the role of internal flux cancellation as a key mechanism for restructuring the magnetic field and driving blowout-jet activity in emerging regions.}
{}

\titlerunning{Flux Cancellation in 3D MHD Flux Emergence}
\maketitle
\nolinenumbers
\section{Introduction}
Magnetic flux cancellation is a phenomenon frequently observed in photospheric magnetogram sequences, and it manifests as
decreases in the magnetic flux when opposite polarity magnetic elements approach each other and interact  (\citealt{1985AuJPh..38..929M}, \citealt{livi1985}, \citealt{1988SoPh..117..243M}). 
It occurs all over the solar disk, including the quiet Sun, active regions, and coronal holes. It is often interpreted as being a consequence of magnetic reconnection that occurs in the low solar atmosphere \citep{yurchyshyn2001}. \cite{zwaan1987} proposed three  scenarios 
to explain magnetic flux cancellation depending on where magnetic reconnection takes place: below, above, or at magnetogram level, giving rise to the emergence of U loops, the submergence of $\mathrm{\Omega}$ loops, or emerging and submerging U or $\mathrm{\Omega}$ loop pairs, respectively. All three cases are well represented in the first figure in the work of \cite{Kubo2007}.

Several observational studies
have shown evidence of either emerging U loops or submerging $\mathrm{\Omega}$ loops in magnetic flux cancellation regions.
\cite{van2000} used magnetograms and dopplergrams from the Michelson Doppler Imager (MDI) and Solar and Heliospheric Observatory (SOHO) and showed direct evidence of the emergence of $\mathrm{U}$ loops. Soon after, \cite{bernasconi2002} studied moving dipolar features (MDFs) using SOHO and Transition Region and
Coronal Explorer (TRACE) data as well as Flare Genesis Experiment (FGE) magnetograms and dopplergrams for an observed active region, and they interpreted them as the photospheric signatures of small-scale U loops produced by reconnection below the surface. \cite{rubio2005} analysed visible and infrared spectropolarimetric observations of flux cancellation in the periphery of sunspots. They detected strong upflows in the cancelling flux consistent with the emergence of U loops. \cite{harvey1999} argued that emerging U loops are less common than submerging $\mathrm{\Omega}$ loops. \citet{Yang} analysed photospheric magnetograms of a coronal hole region and found persistent downflows at locations of cancellation, concluding that the disappearing flux results from the submergence of $\mathrm{\Omega}$ shaped loops, consistent with earlier findings by \cite{Chae2004}. More recently, \cite{takizawa2012} studied MDI data and discovered prominent downflows of $1.5~\mathrm{km\,s^{-1}}$
consistent with the submergence hypothesis. \cite{Kaithakkal2019} analysed SUNRISE/IMaX observations and found downflows at cancellation sites, concluding that small-scale cancellations are driven by the submergence of magnetic flux.

Although several studies have identified the physical mechanism responsible for magnetic flux cancellation, distinguishing between the different cancellation scenarios is not always straightforward.  \cite{Kubo2007} studied observations of 12 collision events of opposite polarities and detected the formation of new coronal loops connecting the cancelling elements, but they were not able to identify $\mathrm{\Omega}$ or $\mathrm{U}$ shaped loops. More recently, \cite{Bura} analysed high-resolution Interface Region Imaging Spectrograph (IRIS), Goode Solar Telescope (GST), and Solar Dynamics Observatory (SDO) observations and found that small-scale magnetic flux cancellation is closely associated with jet formation. They reported that opposite polarity magnetic elements emerge and subsequently cancel at the jet base, suggesting that unresolved magnetic reconnection in the chromosphere drives the observed activity without providing clear evidence of either flux emergence or submergence as the underlying mechanism.

The implications of magnetic flux cancellation are far-reaching for a wide range of solar phenomena. It plays a fundamental role in the formation of pre-eruptive magnetic flux ropes (MFRs) and coronal sigmoids (e.g. \cite{van1989,amari2003,mackay2006,Aulanier2009,green2011,savcheva2012,hassanin2022,patsourakos2025, Furuseth2026}) as well as in triggering eruptive events across multiple scales, from small-scale jets to large-scale coronal mass ejections (e.g. \cite{liu2004,green2011,savcheva2012,panesar2016,yardley2018,panesar2018,chintzoglou2019,moore2022}). It has also been associated with Ellerman bombs (e.g.\cite{Schmieder2002,Pariat2004,reid2016}), IRIS bombs (\cite{peter2014}), quiet Sun activity (e.g. \cite{Iida2010,mou2018,McGlasson2019,nelson2024}), chromospheric flows and flares (e.g. \cite{zhang2001,zuccarello2008,panesar2018,panesar2023,Li2026}), and chromospheric and coronal heating and the acceleration of the solar wind (\cite{Gosic2018,Pontin2023}). Magnetic flux cancellation is further closely linked to the formation of coronal jets and chromospheric jet-like flows (e.g. \cite{mou2018,panesar2018,kumar2019,moore2022,panesar2023,nelson2024}), highlighting its importance in driving dynamic activity in the solar atmosphere. Observational studies support these connections. \cite{reid2016}, using high-resolution Crisp Imaging Spectro-polarimeter and IRIS observations combined with Helioseismic and Magnetic Imager (HMI) magnetograms, showed that Ellerman bombs form at locations where opposite polarity magnetic patches converge and cancel. Similarly, \cite{mou2018}, analysing Atmospheric Imaging Assembly (AIA) and HMI data, found that brightenings and eruptions of quiet-Sun coronal bright points occur during episodes of photospheric flux cancellation at the underlying small bipoles.

A series of magnetohydrodynamic (MHD)
simulations with varying complexity and initial and boundary conditions have been used to study magnetic flux cancellation. 
First, surface-driven models have been used to apply postulated converging and shearing photospheric flows to drive magnetic flux cancellation (e.g.\cite{van1989, amari2003, mackay2006, Aulanier2009, Syntelis2019,
Faerder2023, Pontin2023}). Although some of these studies reported submerging loops, this scenario remains more difficult to detect observationally than emerging U loops. However, many of these studies further explored the role of cancellation in various solar phenomena. \cite{Syntelis2019} used a 2D resistive MHD simulation of two opposite polarities at the photosphere that move towards each other and eventually cancel. They associated this cancellation with the appearance of strong current sheets and the release of nano-flare-scale energy. \cite{Pontin2023} argued that small-scale flux cancellation could be associated with magnetic reconnection in the lower solar atmosphere, which can contribute to chromospheric and coronal heating and supply energy to the solar wind.

Magnetic flux cancellation has also been studied using flux emergence simulations.
In contrast to the surface-driven simulation described in the previous paragraph, 
which requires ad hoc surface shearing and converging flows to drive cancellation, flux emergence
simulations self-consistently generate such flows.
These simulations employ the emergence of flux sheets or twisted flux tubes. 

For flux sheet models, \citet{panesar2023} used Bifrost simulations of a horizontal magnetic flux sheet and found that small-scale jets are closely linked to flux cancellation at photospheric neutral lines. The resulting reconnection in the lower atmosphere drives plasma ejections that share key characteristics with larger-scale coronal jets, including collimated outflows and multi-thermal plasma signatures. These findings support a unified picture in which flux cancellation acts as a fundamental mechanism for jet formation across different spatial scales, in agreement with our results. Similarly, \citet{moreno2025} analysed a cancellation event in a 3D radiative Bifrost simulation where an emerging bipole interacts with pre-existing opposite polarity. In this case, magnetic reconnection occurs at chromospheric heights, and the submergence of post-reconnection loops leads to flux cancellation, which is also linked to the formation of flux ropes. Additionally, \citet{Chen2026} performed a 3D radiative MHD simulation using the MURaM code in which magneto-convective motions self-consistently generate a twisted magnetic structure that emerges through the photosphere. This emerging twisted flux interacts with the ambient open magnetic field, leading to magnetic reconnection and the onset of a blowout jet. The evolution is accompanied by flux cancellation at the base of the jet as well as the transfer of twist from closed to open field lines, driving untwisting motions.

In contrast, flux tube simulations typically adopt twisted initial magnetic structures, most often with a uniformly distributed twist. \citet{Magara2011} showed that magnetic flux cancellation arises from the evolution of U-shaped magnetic loops at the photospheric polarity inversion line (PIL). Shallow U loops can rise as plasma drains from their dips, whereas deeper U loops are unable to emerge and instead undergo reconnection, leading to flux cancellation signatures. Earlier studies by \citet{Magara2006} and \citet{Cheung2008} employed the emergence of uniformly twisted magnetic flux tubes and considered different initial twist amplitudes. Both studies found that magnetic flux cancellation is associated with strong downflows at the photosphere and occurs preferentially along PILs. In \citet{Cheung2008}, cancellation is primarily driven by convective motions, which bring opposite polarities together and lead to the formation of current layers and magnetic reconnection at low atmospheric heights. This process produces small-scale loops that submerge below the photosphere, resulting in apparent flux cancellation without being directly linked to energetic coronal phenomena. In contrast, \citet{Magara2006} demonstrated that flux cancellation can be part of a more dynamic magnetic evolution where the interaction of emerging twisted fields leads to magnetic reconnection and plasma ejections in the corona, indicating that cancellation may be associated with energetic events such as jet-like outflows.
More recently, \citet{patsourakos2025} used a uniformly twisted emerging flux tube in Lare3D and reported flux cancellation between the magnetic tails of the emerging bipolar region. They associated this process with the rise of U loops formed through reconnection along the internal PIL and with the formation of a magnetic flux rope that later erupted as a blowout jet. However, they found that only about $\mathrm{1\%}$ of the total emerging flux participates in the cancellation process, much lower than observational estimates.

In this work, we performed a 3D MHD simulation in order to study flux cancellation due to the emergence of a twisted horizontal flux tube from the convection zone to the photosphere. As discussed above, a few numerical studies have focused on detecting and studying cancellation events that can occur naturally from the emergence of twisted flux tubes. We followed the work of \cite{Murray2008} and employed a non-uniformly twisted flux tube in order to produce a complex active region in which magnetic elements could cancel. In Section \ref{model}, we present the numerical setup and the initial state of the flux tube. In Section \ref{results}, we describe the general evolution and the main flux cancellation event. In Section \ref{4}, we interpret the cancellation process and examine its association with flux rope formation and the blowout jet. We summarise our results in Section \ref{5} and present two secondary cancellation events in Appendix \ref{3.5}.

\section{Model}
\label{model}

\subsection{Setup}

We solved the three-dimensional time-dependent resistive magnetohydrodynamic (MHD) equations in Cartesian geometry using the Lare3D code \citep{Arber2001}. The background atmosphere was assumed to be in hydrostatic equilibrium. A uniform oblique ambient magnetic field was included, inclined by $\mathrm{\theta = \mathrm{11^\circ}}$ with respect to the vertical ($\mathrm{z}$-axis) and with an azimuthal angle of $\mathrm{\phi = \mathrm{183^\circ}}$ in the horizontal ($\mathrm{x}$--$\mathrm{y}$ plane). Its strength was set to $\mathrm{B_{\mathrm{amb} }\simeq 0.033}$ in code units (corresponding to $\approx \mathrm{10}\,\mathrm{G}$). The openness of the field lines gives the region of interest the characteristics of a coronal-hole-like environment. A horizontal twisted magnetic flux tube was embedded in the upper convection zone. Both viscous heating and Joule heating were included, and a uniform finite dimensionless value of $\mathrm{\eta =10^{-2}}$ was applied for the resistivity.
\begin{figure}[h]
        \hspace{1cm}
    \includegraphics[width=7cm]{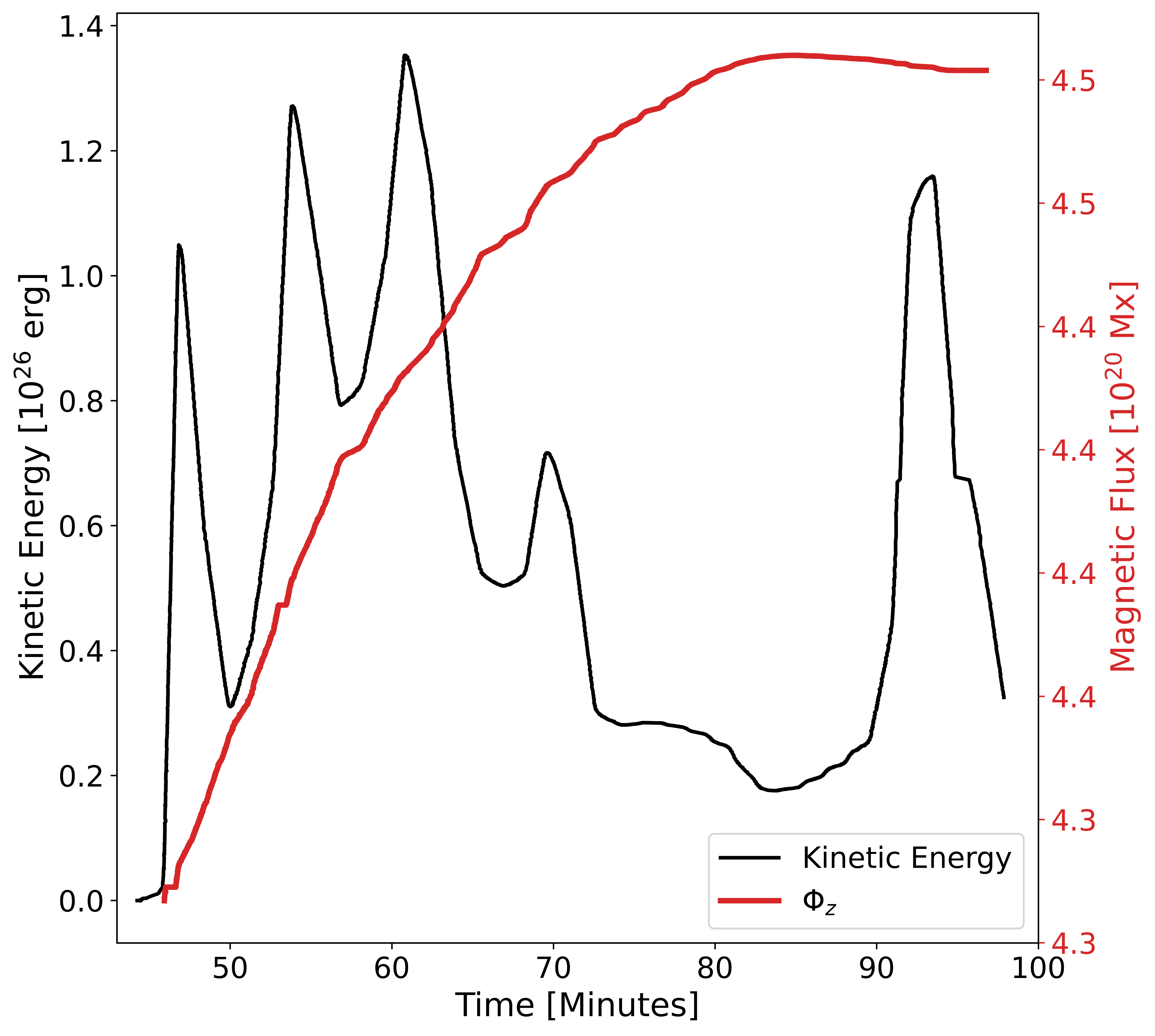} 
    \caption{Temporal evolution of the kinetic energy in the corona (black line) and of the net unsigned magnetic flux in the photosphere (red line). }
\label{KE}
\end{figure}
The computational domain has a physical size of $\mathrm{64.8 \times 64.8 \times 64.8\,\mathrm{Mm^3}}$ in the $\mathrm{(x,y,z)}$ directions, and is covered by a uniform grid of $\mathrm{420 \times 420 \times 420}$ cubic elements. The flux tube was initially placed at a depth of $\mathrm{z_0 = -2.1\,\mathrm{Mm}}$ below the photosphere, with its axis aligned with the $y$-axis. The simulation output had a temporal cadence of $\mathrm{25.6}\,\mathrm{s}$ between consecutive snapshots. The axial magnetic field of the flux tube is given by
$\mathrm{B_y = B_0 \exp\left(-\frac{r^2}{R^2}\right)}$
where $\mathrm{B_0 = 3900}\,\mathrm{G}$ is the peak field strength and $\mathrm{R = 0.45\,\mathrm{Mm}}$ is the tube radius. The azimuthal component of the tube field is
$\mathrm{B_\theta = \alpha r B_y}$
where $\alpha$ is the twist parameter and $\mathrm{r^2 = x^2 + (z - z_0)^2}$ is the radial distance from the tube axis. To initiate the flux tube emergence, a density deficit along the tube axis was imposed, making its central portion buoyant. The deficit is proportional to the local pressure excess inside the tube and is modulated along the axial direction by a Gaussian profile with length scale $\mathrm{\lambda=0.9}\,\mathrm{Mm}$.

\subsection{Initial state of the tube and the role of twist}
\label{initial}

The numerical simulation considers an emerging flux tube with a radially varying twist profile at $\mathrm{t=0}$. The non-uniform twist profile was originally introduced by \citet{Murray2008}:

    \begin{equation}
    \alpha_{inc}(r) = c\left(1 - \exp\left(-\frac{r^2}{k^2}\right)\right).
    \end{equation}

The constant $\mathrm{k=5}$ controls the rate of increase in twist away from the centre of the tube, while $\mathrm{c=0.4}$ sets the amplitude of the twist profile. Under this assumption, the inner field lines remain untwisted, while the twist gradually increases away from the tube axis. The majority of the twisted field lines are located at the edge of the flux tube, which in total has an area-weighted average over the tube cross-section of $\mathrm{\langle \alpha_{inc} \rangle = 0.09}$. Our setup is similar to that of \cite{patsourakos2025}. The main differences, in addition to the twist treatment, are the inclusion of partial ionisation effects, a deeper initial flux tube position, and a different initial magnetic field strength.

\begin{figure*}
\centering
\begin{minipage}[c]{13cm}
    \includegraphics[width=12cm]{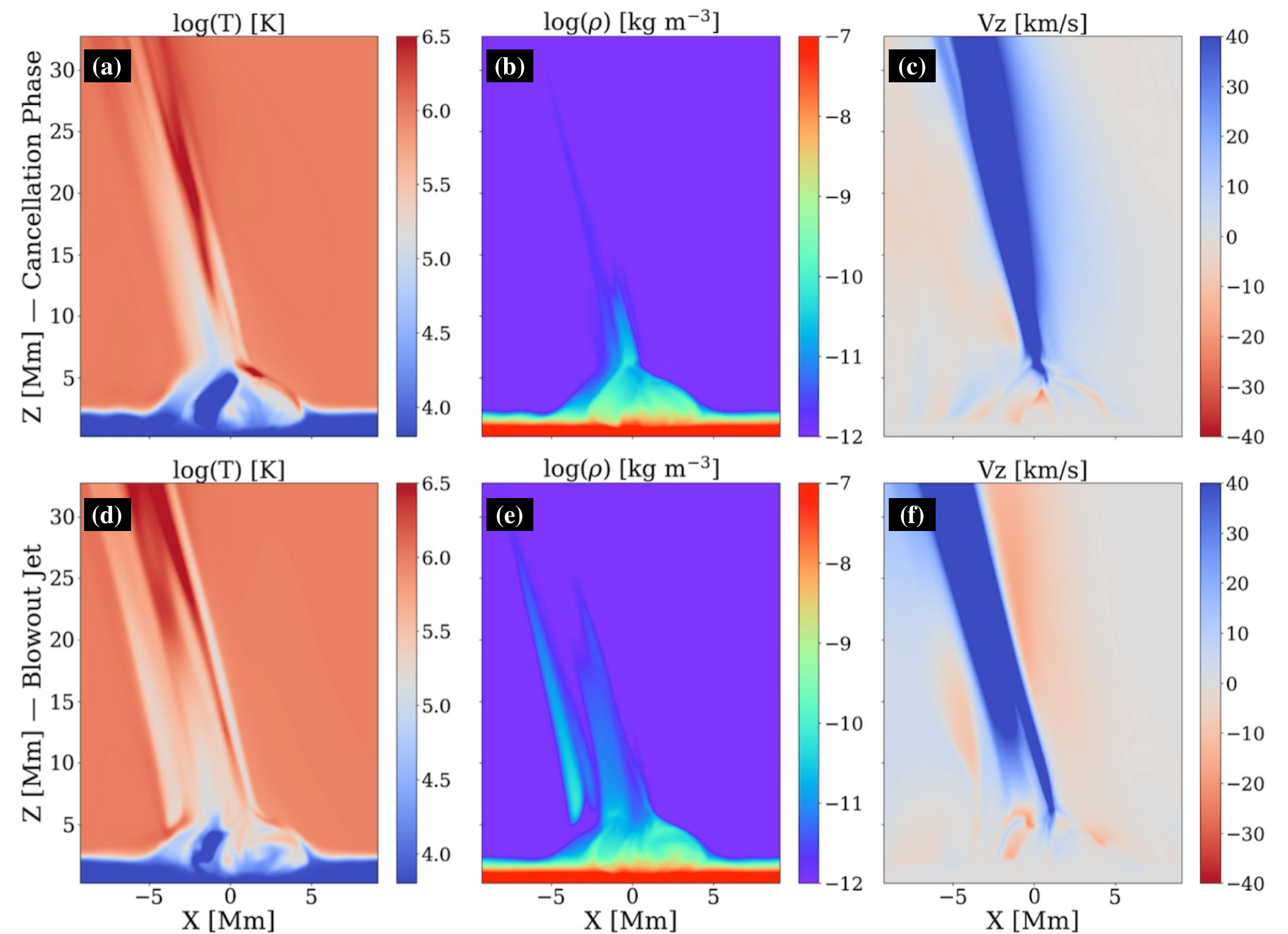}
\end{minipage}\hfill
\begin{minipage}[c]{5.4cm}
\caption{
Two-dimensional distributions in the xz plane of temperature,
plasma density, and vertical velocity (from left to right) at $\mathrm{y=-0.7~Mm}$. The upper row shows the standard jet at $\mathrm{t\approx51~min}$, while the lower row corresponds to the blowout jet at $\mathrm{t\approx64~min}$. The standard jet exhibits a narrow collimated spire, whereas the blowout jet is broader, hotter, and more spatially extended, with stronger upward plasma motions.
}
    \label{multi}
\end{minipage}
\end{figure*}

The assumption of a non-uniform initial twist provides a more realistic description of emerging magnetic fields, since uniformly twisted flux tubes represent an idealised case. Several studies have shown how non-uniform twist distributions can naturally arise. For example, \cite{2000ApJ...540..548A,2003ApJ...582..475A} performed 3D ideal MHD simulations of emerging flux tubes and found that adiabatic expansion, vorticity, shear motions, and reconnection can significantly modify the twist, producing a strongly twisted axial core compared to the outer layers. Using 3D simulations, \cite{fan2009} showed that as the flux rope expands into the low density atmosphere, the twist per unit length decreases in the emerged part, while the subsurface part remains more strongly twisted. In addition, \cite{hotta2012} demonstrated that twist can be generated self-consistently through flows at the base of the convection zone. Finally, \cite{toriumii2011,toriumi2012} showed that during the rise and expansion near the photosphere, the twist becomes concentrated around the core and decreases outward, naturally producing a non-uniform distribution.
In our simulation, the magnetic flux tube starts as a weakly twisted structure, with an initial axial magnetic flux of $\mathrm{3.4 \times 10^{19}}$ $\mathrm{Mx}$. Still, it successfully emerges into the solar atmosphere, and reconnects with the ambient field, creating jets. Thus, even low initial twist can produce energetic phenomena. Twist is a key parameter that strongly affects emerging flux tubes. \cite{Magara2001, Magara2006} showed that strongly twisted flux tubes concentrate flux around the polarity inversion line, while weakly twisted flux tubes have more scattered flux and emerge faster (e.g. \citealt{Moreno1992}, \citealt{Moreno1996}, \citealt{SyntelisREC}, \citealt{Luoni2011}). Similarly, \cite{Murray2006} found that weakly twisted fields can form magnetic side lobes, while \cite{Sykora2008} showed that stronger twist leads to a more coherent emergence and enhanced current concentrations. In general, several studies (\cite{Toriumi2011, Archontis2010, 2012A&A...537A..62A, Syntelis2019, Zhuleku, Alissandrakis_2025, Karantanis2026}) have shown that the initial twist and magnetic field strength, as well as the initial geometry of the field can influence the emergence rate and horizontal expansion of flux tubes.

\section{Results and discussion}
\label{results}
\setlength{\parskip}{0.1pt}
\setlength{\parindent}{15pt}

\subsection{General evolution of the simulation}

\label{3.1}

Our simulation has a total duration of $\mathrm{\approx 100}$ $\mathrm{minutes}$. During this time interval, the flux tube's buoyant part undergoes coherent emergence through the photosphere, beginning at $\mathrm{\approx 40}$ $\mathrm{minutes}$ in the simulation. To obtain some context of the simulation, especially in the photosphere and above it, we plotted the kinetic energy above the photosphere ($\mathrm{z>5}$ $\mathrm{Mm}$), as well as the magnetic flux through the photospheric plane as a function of time (Fig.   \ref{KE}).

Fig.   \ref{KE} shows that the kinetic energy exhibits multiple spikes that correspond to a series of energetic events. The first peak of the coronal kinetic energy, at $\mathrm{t} \mathrm{\approx 46}$ $\mathrm{minutes}$, is associated with the ejection of a small plasmoid from low heights ($\mathrm{z} \mathrm{\approx 1}$ $\mathrm{Mm}$) into the upper corona. This structure consists of tangled magnetic field lines that escape from near a reconnection site in  an off-centre location on the photospheric plane ($\mathrm{x \approx 0.5}$ $\mathrm{Mm}$, $\mathrm{y} \approx 1.4$ $\mathrm{Mm}$) and heats the surrounding plasma up to $\mathrm{T} \approx 5 \times 10^{5}$ $\mathrm{K}$, but is not related to any flux cancellation event. 

The next peak of the kinetic energy at $\mathrm{t=51~minutes}$ is more pronounced and corresponds to a standard jet, while the following peak at $\mathrm{t=63~minutes}$ corresponds to a blowout jet. {The distinction between standard and blowout jets was introduced by \cite{moore2010}, who classified them primarily by jet morphology: standard jets exhibit a relatively narrow spire, whereas blowout jets develop a substantially broader spire, a distinction that we quantify later in this section for the two events in our simulation. The broader spire is associated with a more extensive opening of the overlying closed magnetic structure and a broader, coronal mass ejection-like eruption.} Using SDO/AIA and HMI observations, \cite{Liu2011} showed that interchange reconnection occurs over a dome-like base during the standard phase, whereas during the blowout phase the closed dome opens and the erupting magnetic structure expands laterally into the corona. \cite{sterling2022} further emphasised that both types of jets involve internal reconnection beneath the rising erupting mini-filament and external reconnection between the closed field and the ambient open field.

Numerical studies have provided further insight into the physical and topological differences between the two jet types. 
\cite{Archontis2013} used 3D MHD flux emergence simulations to study the transition from the standard to the blowout phase, while \cite{moraitis2024} found that the onset of blowout jets is accompanied by a decrease in both free magnetic energy and current-carrying helicity, the latter serving as a proxy for magnetic twist. 
This decrease indicates that the enhanced dynamics of the blowout phase are associated with the release of previously stored magnetic energy and helicity, consistent with observational estimates of solar jets \citep{Nindos2024}. 

{In our simulation, the key topological difference between the two events is that a coherent magnetic flux rope has not yet formed during the standard jet, whereas an MFR is present during the blowout jet. During the standard jet, external reconnection between the envelope of the emerging magnetic field and the ambient field produces the jet, while the core of the closed emerging structure remains largely confined. In contrast, the blowout jet occurs after the formation of the MFR through flux cancellation. During the blowout phase, the MFR expands and undergoes a partial eruption: part of the confining closed field reconnects with the ambient field and opens, transferring twisted magnetic flux from the MFR to the ambient open field, while a substantial part of the MFR core remains in the lower corona. Therefore, the distinction between the two events does not require the complete ejection of the MFR. Instead, the presence and partial eruption of the MFR, together with the opening of the confining field and the transfer of twisted flux to the ambient open field, provide the physical and topological signatures of the blowout phase. The evolution of the open and closed twisted magnetic flux shown in Sect.~4.3 and Fig.~12 provides quantitative evidence of this topological transition.}

\begin{figure*}[htbp]
    \hspace{-0.1cm}        \includegraphics[width=18cm]{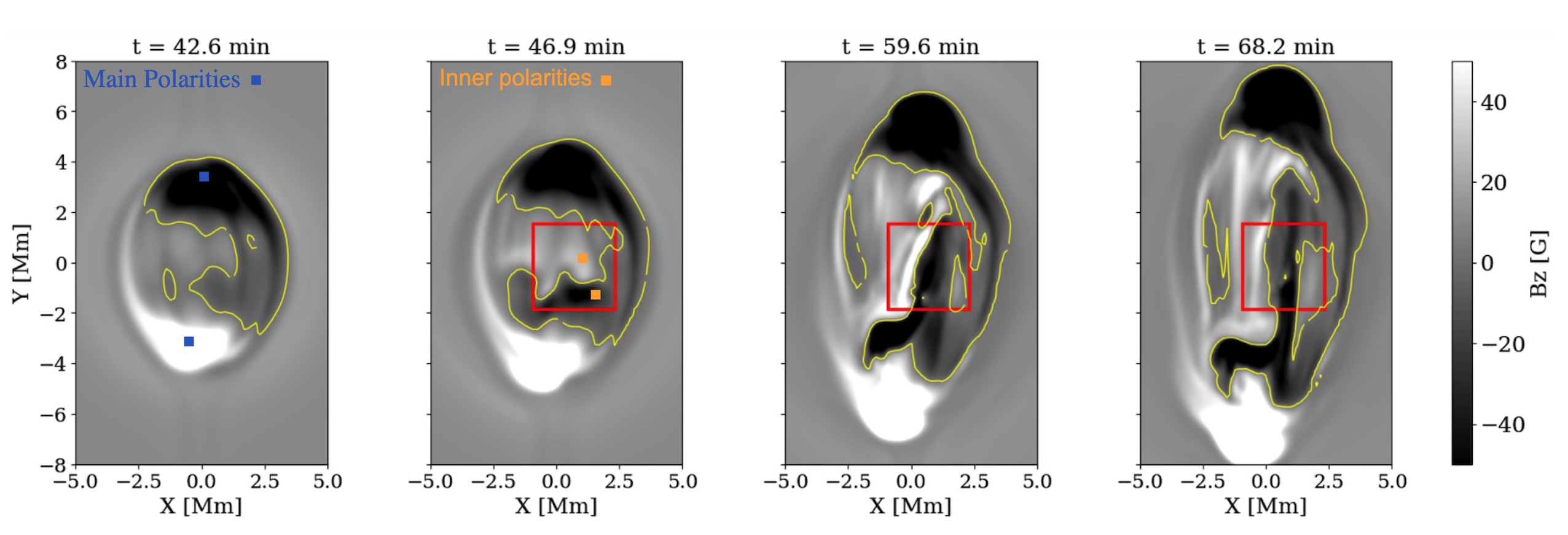} 
    \caption{Distribution of $\mathrm{B_z}$ at $\mathrm{z}=0.05$ $\mathrm{Mm}$ for four snapshots during our simulation. Panels from left to right correspond to  $\mathrm{t\approx 43, 47, 60, 68}$ minutes. Black regions indicate negative polarities, while white regions denote positive polarities. The yellow line indicates the PIL. The red box encloses the region where the main cancellation event of our simulation takes place. Blue and orange squares mark the outer and inner polarities, respectively. An associated movie is available online. }
\label{bz}
\end{figure*}

We focus mainly on these two jets, as they occur during the major flux cancellation episode at the photospheric plane which will be discussed later. For the rest of the simulation, we observe several smaller peaks that correspond to eruptive jets. {At $\mathrm{t\approx90~min}$, we detect another peak in the coronal kinetic energy, which corresponds to a secondary blowout jet. Following the primary eruption, the relaxed magnetic system continues to expand, accompanied by further magnetic reconnection between the reconfigured emerging field and the inclined ambient field. This event is directly linked to the main flux cancellation episode, as the initial eruption provided the necessary magnetic reconfiguration for this late activity. Similar to the initial eruption, hot material from the lower atmosphere is again accelerated upwards, although we do not study this secondary blowout jet further.} Fig.  \ref{KE} shows the temporal evolution of the net unsigned photospheric magnetic flux. The flux is initially small and starts to increase rapidly at $\mathrm{t} \approx 45$ $\mathrm{minutes}$ until $\mathrm{t} \approx 80$ $\mathrm{minutes}$, due to the emergence of the magnetic field in the upper photosphere. Thereafter, the flux reaches a plateau as most of the emergence has been completed and a substantial fraction of the emerged field has reached the lower corona.

  Fig. \ref{multi} shows the 2D distributions of temperature, density, and vertical velocity (from left to right) during the standard  (upper row) and  another during the onset of the blowout jet (lower row). These slices are taken at $\mathrm{y=-0.7}$ $\mathrm{Mm}$, rather than at the centre of the domain, because the main activity is shifted away from the central plane. {Following the morphological framework described above, we estimated the spire widths from the horizontal extent of the density-enhanced jet material along the $x$ direction in the $x$--$z$ plane shown in Fig.~2. At a representative coronal height of $\mathrm{z=15~Mm}$, the spire widths of the first event at $\mathrm{t=51~min}$ and the second event at $\mathrm{t=63}$--$\mathrm{64~min}$ are approximately $\mathrm{0.8~Mm}$ and $\mathrm{5.1~Mm}$, respectively. Thus, the spire of the second event is about six times broader than that of the first. The width of the underlying loop base during the second event, estimated from the horizontal extent of the density-enhanced loop system over $\mathrm{z=3.5}$--$\mathrm{5.0~Mm}$, is approximately $\mathrm{7.2~Mm}$. Therefore, the spire of the second event reaches about $\mathrm{70\%}$ of the loop-base width and is comparable to the base in scale, although it remains somewhat narrower. Based on this morphological comparison, we classify the first event as a standard jet and the second event as a blowout jet.}

The upper row of Fig.~2 corresponds to $\mathrm{t\approx51}$ $\mathrm{min}$, when reconnection between the emerging front and the magnetised atmosphere has already begun and the jet spire is aligned with the ambient field. The standard jet lasts for approximately $\mathrm{9~min}$ and heats the plasma up to a maximum temperature of approximately $\mathrm{6.2}$ $\mathrm{MK}$. The spire is multithermal, with a small cooler surge on its left side at approximately $\mathrm{10^{4.5}}$ $\mathrm{K}$. The jet has typical coronal densities of  $\mathrm{\approx10^{-10}-10^{-9}}$ kg m$^{-3}$, with the densest material located below $\mathrm{z=10}$ $\mathrm{Mm}$. Moreover, its mean vertical velocity reaches approximately $\mathrm{160}$ km s$^{-1}$. The lower row of Fig.~2 shows the corresponding distributions during the onset of the blowout jet. This jet lasts for approximately 12 min and produces a stronger and more spatially extended plasma response. {Considering cells with $\mathrm{v_z>50}$ km s$^{-1}$ within the selected three-dimensional jet volume, the mean temperature increases from approximately $\mathrm{1.41}$ $\mathrm{MK}$ in the standard jet to $\mathrm{1.74}$ $\mathrm{MK}$ in the blowout jet, corresponding to an increase of approximately $\mathrm{23\%}$.} The maximum temperature reaches approximately $\mathrm{10}$ $\mathrm{MK}$, about $\mathrm{60\%}$ higher than the maximum temperature of $\mathrm{6.2}$ $\mathrm{MK}$ reached during the standard jet. Regarding the vertical velocity, the blowout jet exhibits large upflows with maximum values of approximately $\mathrm{221}$ km s$^{-1}$, while its mean vertical velocity is approximately 15--25\% greater than that of the standard jet. Several smaller and localised downflows are also present, particularly around the twisted structure at $\mathrm{z\approx2}$ Mm. This structure and its associated current concentrations are directly connected with the photospheric flux cancellation episode discussed in the next section.

\subsection{Main flux cancellation event}
\label{3.2}

In this section, we focus on the first and largest flux cancellation event detected in our simulation. This is the most notable cancellation event of our simulation given its duration, its spatial extent relative to the total emerging region, and its impact on the overlying atmosphere. In the Appendix, we also discuss two smaller partial flux cancellation events detected during the simulation. 

\begin{figure}[htbp]
\hspace{1.1cm}
    \includegraphics[width=6.5cm]{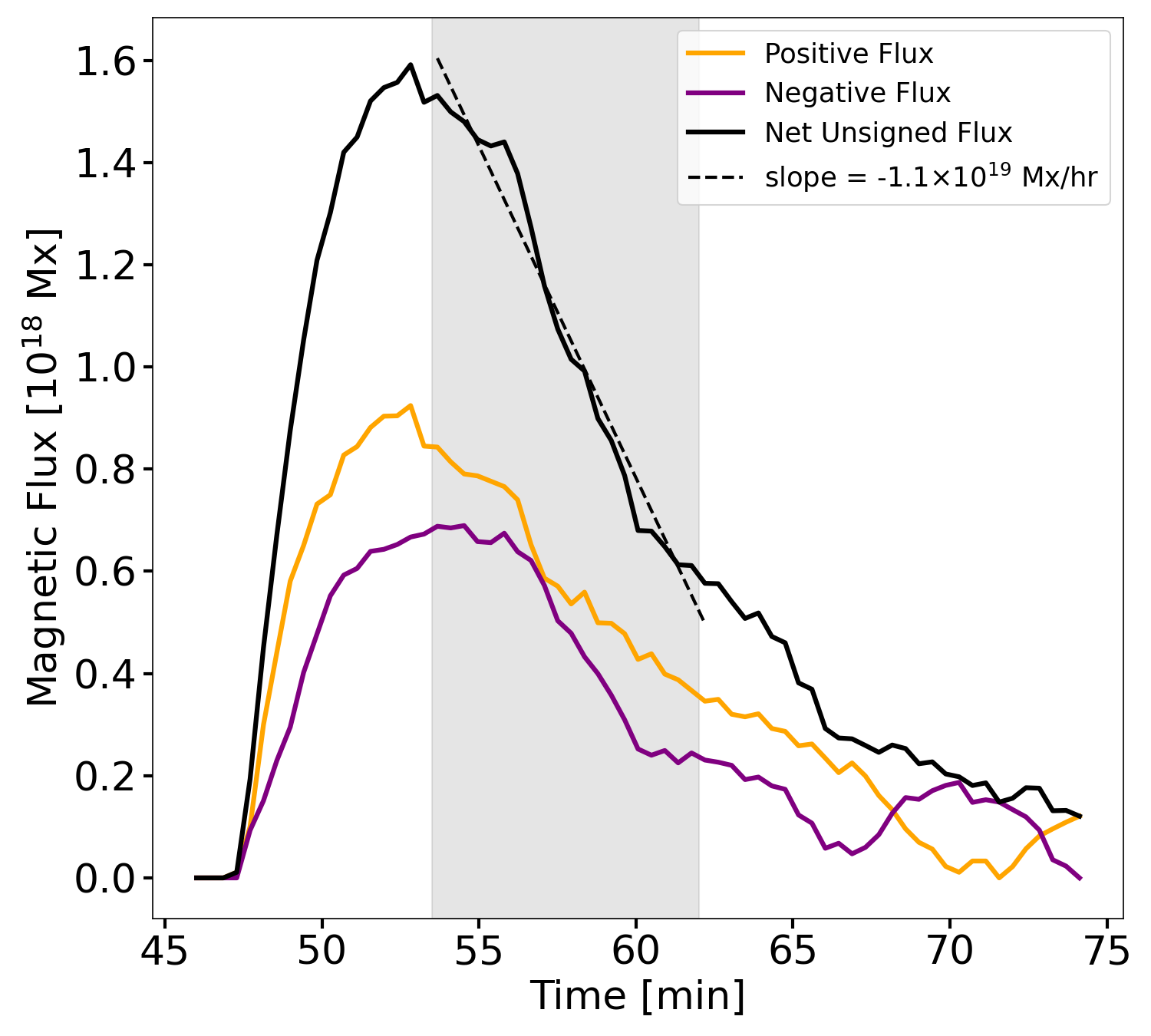}
    \caption{Temporal evolution of the net unsigned magnetic flux (black), positive
flux (orange), and absolute negative flux (purple) in the red box shown
in Fig. 3. The grey shaded region marks the main cancellation interval.
The black dashed line shows a linear fit to the net unsigned flux.}
   \label{fluxcan}
\end{figure}

Figure~\ref{bz} shows magnetograms of $\mathrm{B_z}$ at $\mathrm{z = 0.05}$ $\mathrm{Mm}$ for selected time steps. At $\mathrm{t \approx 43}$ $\mathrm{minutes}$, we observe the  emergence of the outer magnetic polarities. Blue squares mark the main outer polarities, while orange squares
mark the inner opposite-polarity elements involved in the cancellation. The yellow lines show the location of the photospheric polarity inversion lines (PILs), where $\mathrm{B_z}$ changes sign. Hereafter, PIL refers to the polarity inversion line of the smaller inner bipole within the red box. The configuration resembles the typical pattern of an emerging bipole; however, the magnetic tails appear relatively extended and weak, remaining well separated without significant interaction during the simulation. This behaviour can be attributed to the  non-uniform and overall weak twist of the emerging flux tube, which results in a limited injection of twist into the lower atmosphere. This behaviour contrasts with simulations of flux emergence with constant or stronger twist, where the magnetic tails appear more pronounced, elongated, and coherent (\cite{patsourakos2025}).

\begin{figure*}
\centering
\begin{minipage}[c]{12cm}
    \includegraphics[width=12cm]{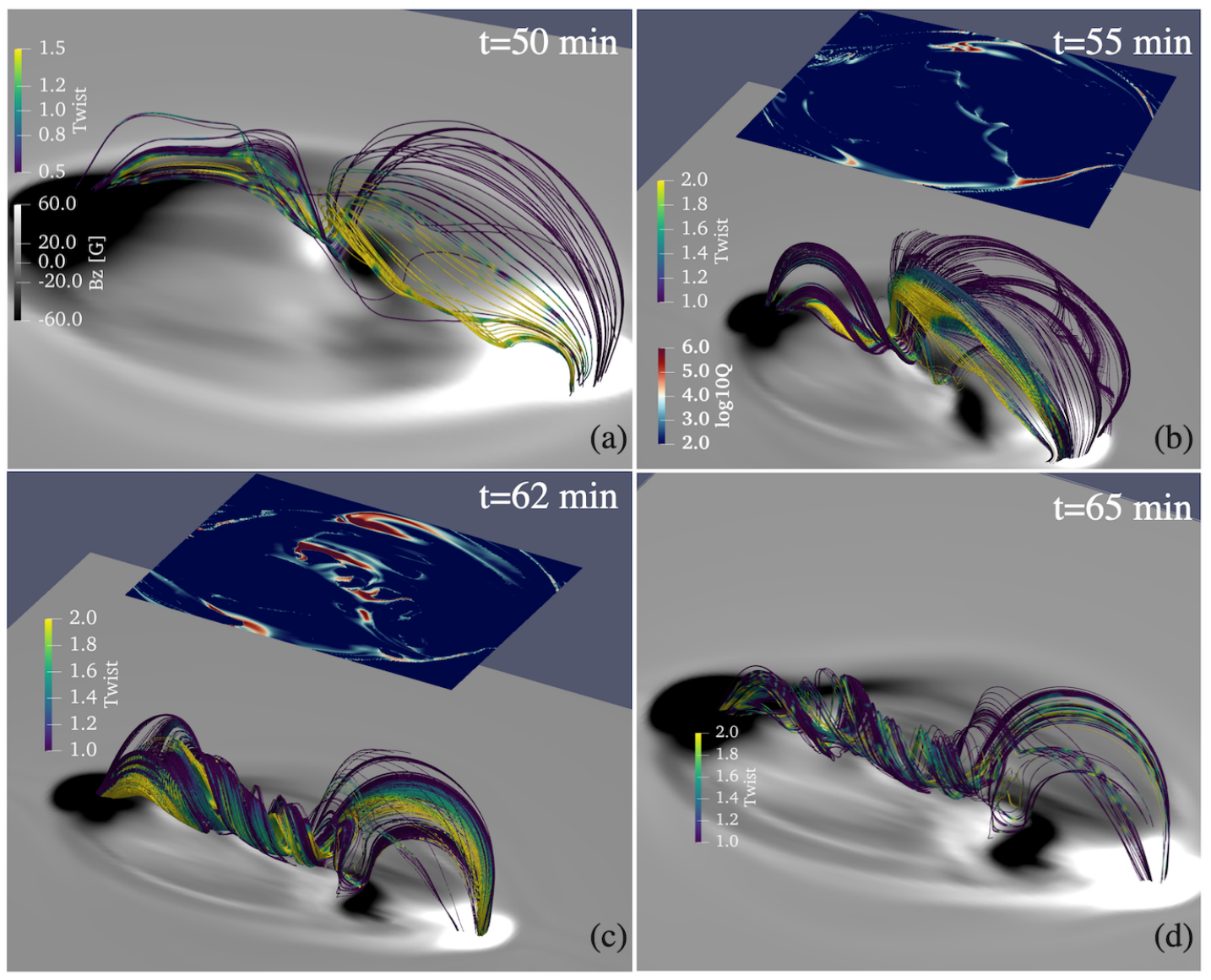}
\end{minipage}\hfill
\begin{minipage}[c]{5.3cm}
\caption{Three-dimensional view of the magnetic field close to the photospheric level at $t \approx 50$, 55, 62, and 65 min, corresponding to (a) the pre-cancellation phase, (b) the onset of flux cancellation, (c) the late cancellation phase and formation of the twisted magnetic structure, and (d) the post-cancellation phase, during the onset of the blowout jet. The closed field lines are coloured according to their twist, with yellow and purple indicating higher and lower twist values, respectively. The greyscale background shows the photospheric $B_z$ distribution. Maps of $\log Q$ at $z \approx 0$ Mm are shown in panels (b) and (c), highlighting regions of strong gradients in magnetic connectivity. The associated movie is available online.}
    \label{lines1}
\end{minipage}
\end{figure*}
The region where the main flux cancellation event occurs is enclosed by the red box for clarity, and shown in the subsequent panels of Fig. ~\ref{bz}. At $\mathrm{t \approx 51}$ minutes, the red box contains two smaller magnetic elements of opposite polarity (orange squares), which first emerge at $\mathrm{t \approx 46}$--$\mathrm{47}$ minutes, and give rise to  a secondary dipolar structure embedded within the larger pre-existing bipole. During this phase, the negative polarity patch (black) moves towards $+\mathrm{y}$ direction, while the positive polarity patch (white) moves downwards (towards $-\mathrm{y}$). Simultaneously, the pair undergoes a counterclockwise rotation with time, behaving as a coherent system. As the evolution proceeds, their PIL becomes increasingly aligned with the $\mathrm{y}$-axis, and the two patches converge and  begin to interact. In the third panel, the two magnetic elements remain visible and become more elongated. In the last panel ($\mathrm{t = 68}$ $\mathrm{minutes}$), the two magnetic patches have nearly disappeared, remaining only as faint, oppositely signed $\mathrm{B_z}$ stripes that have drifted further apart.

Therefore, from the discussion of  Fig. ~\ref{bz} there are strong indications
of magnetic flux cancellation between the two inner emerging polarities. To quantify and characterise the postulated flux cancellation event, we proceed
with the calculation of the evolution of the magnetic flux inside the red box
of  Fig. ~\ref{bz}. Before presenting the resulting time series, two important 
pre-processing steps are required. First, in order to isolate the contribution of the two cancelling magnetic elements, we retain only pixels with strong vertical magnetic field by imposing a threshold of $\mathrm{|B_z| > 40}~\mathrm{G}$, which is slightly above the standard deviation ($\mathrm{\sim 1\sigma}$) of the $\mathrm{|B_z|}$ distribution within the selected region. This removes weak background field and ensures that the analysis focuses on strong coherent flux concentrations associated with the opposite polarity patches. Second, because the magnetic flux is measured within the fixed red box of Fig. ~\ref{bz}, flux crossing the box boundaries between consecutive snapshots must be excluded to avoid non-physical contributions to the measured fluxes.

In our calculation of magnetic fluxes in the red box
of Fig. ~\ref{bz}, we ignore pixels where the horizontal magnetic footpoint velocity exceeds $0.7~\mathrm{km\,s^{-1}}$. This threshold is motivated by the spatial resolution of the simulation and the output cadence: the grid spacing is approximately $\mathrm{160}~\mathrm{km}$, implying that flows faster than $\mathrm{0.7}~\mathrm{km\,s^{-1}}$ would displace magnetic structures by a significant fraction of a grid cell within only a few time steps. In order to determine the horizontal magnetic footpoint velocities, we followed the method of \citet{demoulin2003}, which accounts for the influence of flux emergence on the horizontal speed  of magnetic footpoints. According to their formulation, the effective footpoint velocity is given by

\begin{equation}
\mathbf{u} = \mathbf{v}_h - \frac{v_n}{B_n}\,\mathbf{B}_h,
\end{equation}

where $\mathrm{v_h}$ is the horizontal plasma velocity at the photospheric plane, 
$\mathrm{v_n}$ is the vertical velocity ($\mathrm{v_z}$), $\mathrm{B_n}$ is the vertical component of the field ($\mathrm{B_z}$), and finally $\mathrm{B_h}$ is the horizontal component of the  
magnetic field in the photospheric plane.
Indeed, we found horizontal footpoint speeds within the red box
with magnitudes in the range $\mathrm{u_x, u_y=[0.20, 3]}$ $\mathrm{km\,s^{-1}}$, consistent
with other studies (\cite{toriumi2012, rempel2014}). 

Figure ~\ref{fluxcan} shows the temporal evolution of the net unsigned flux (black), the positive vertical flux (orange), and the absolute value of the negative flux (purple), calculated within the red box of Fig. ~\ref{bz} at $\mathrm{z \approx 0~Mm}$, over an $\approx$30-minute interval, and excluding pixels that satisfy the two criteria discussed in the previous paragraphs. Initially, all three curves increase due to the emergence of the small magnetic inner bipole. Shortly afterward, at $\mathrm{t \approx 52}$ minutes, all fluxes begin to decrease simultaneously and flux cancellation sets in. The grey vertical band marks the time interval over which the cancellation occurs. A clear indication of genuine flux cancellation is the close balance between the positive and negative flux curves during the decay phase. In our case, the flux remains well balanced, with a total balance ranging from $\mathrm{\sim88 \%}$ to $\mathrm{91 \%}$, indicating that flux cancellation is the dominant process in the red box, despite a small degree of asymmetry. The flux cancellation rate is estimated to be $\mathrm{(1.1 \pm 0.004)\times10^{19}} \mathrm{Mx\,hr^{-1}}$, as derived from a linear fit applied over the interval of significant decline encapsulated by the grey shaded box of Fig. ~\ref{fluxcan}. Integrating over the event, the total cancelled flux amounts to 
$\mathrm{8.2\times 10^{17}}\ \mathrm{Mx}$. This corresponds to approximately $\mathrm{10 \%}$ of the total magnetic flux contained in the inner portion of the region enclosed by the red box, representing a substantial fraction of its magnetic content. Later time steps ($\mathrm{t>75}$ $\mathrm{minutes}$) are excluded from this analysis, despite the continued flux balance, because the magnetic strength of the patches decreases  significantly, rendering our tracking criteria less reliable. Overall, the analysed event exhibits the key signatures of magnetic flux cancellation, including the convergence of opposite polarities in a PIL and a synchronised 
decrease of nearly balanced positive and negative fluxes.

\section{Physical interpretation and consequences of flux cancellation}
\label{4}

\subsection{Interpretation of the flux cancellation process}
\label{4.1}

\begin{figure}[htbp]
\hspace{0.1cm}
    \includegraphics[width=8cm]{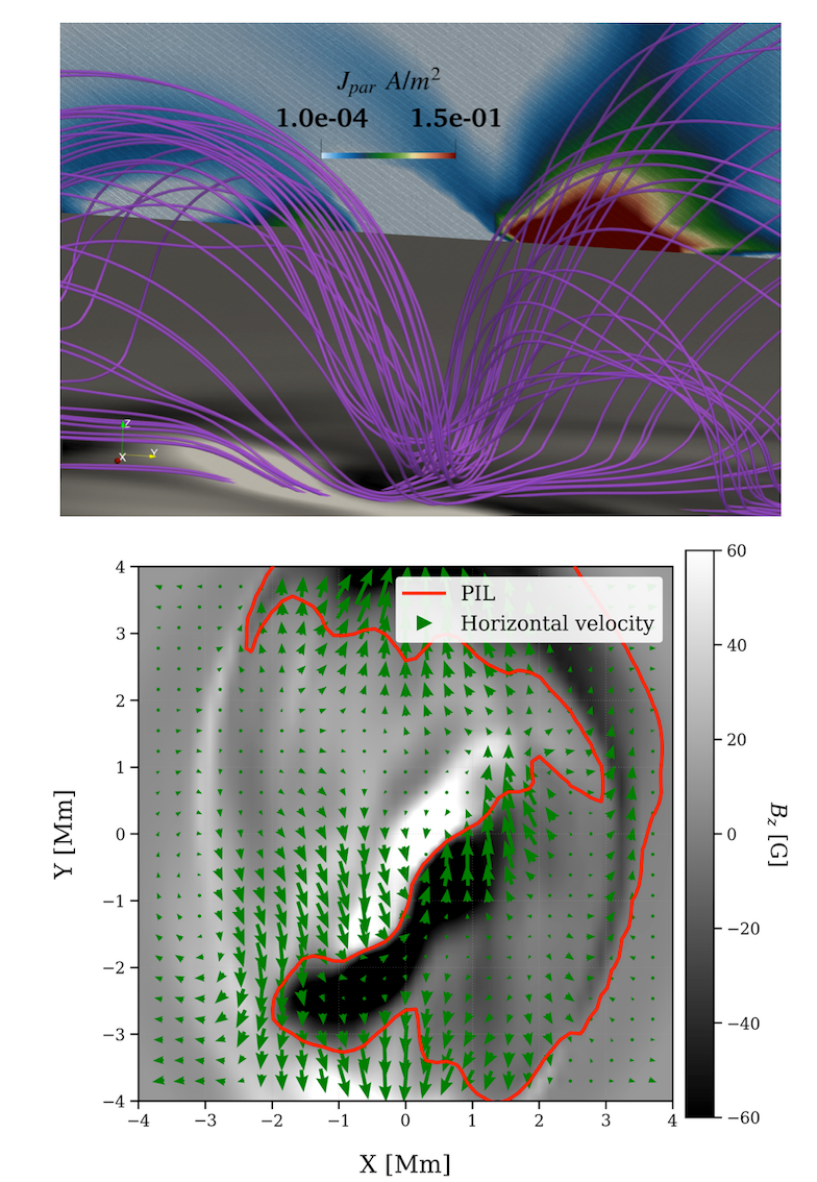}
\caption{Top panel: Side view of the U loops at the beginning of the cancellation process at $\mathrm{t\approx52~min}$. In the background is the map of the current parallel to the magnetic field, at the centre of the loops.
{Bottom panel: Top view of the polarities during the cancellation phase at $\mathrm{t\approx55~min}$. The green arrows correspond to the horizontal velocity of the footpoints close to the photospheric level. The red solid line corresponds to the PIL.} }
   \label{uloops}
\end{figure}

\begin{figure}
    \centering
    \includegraphics[width=7.0cm]{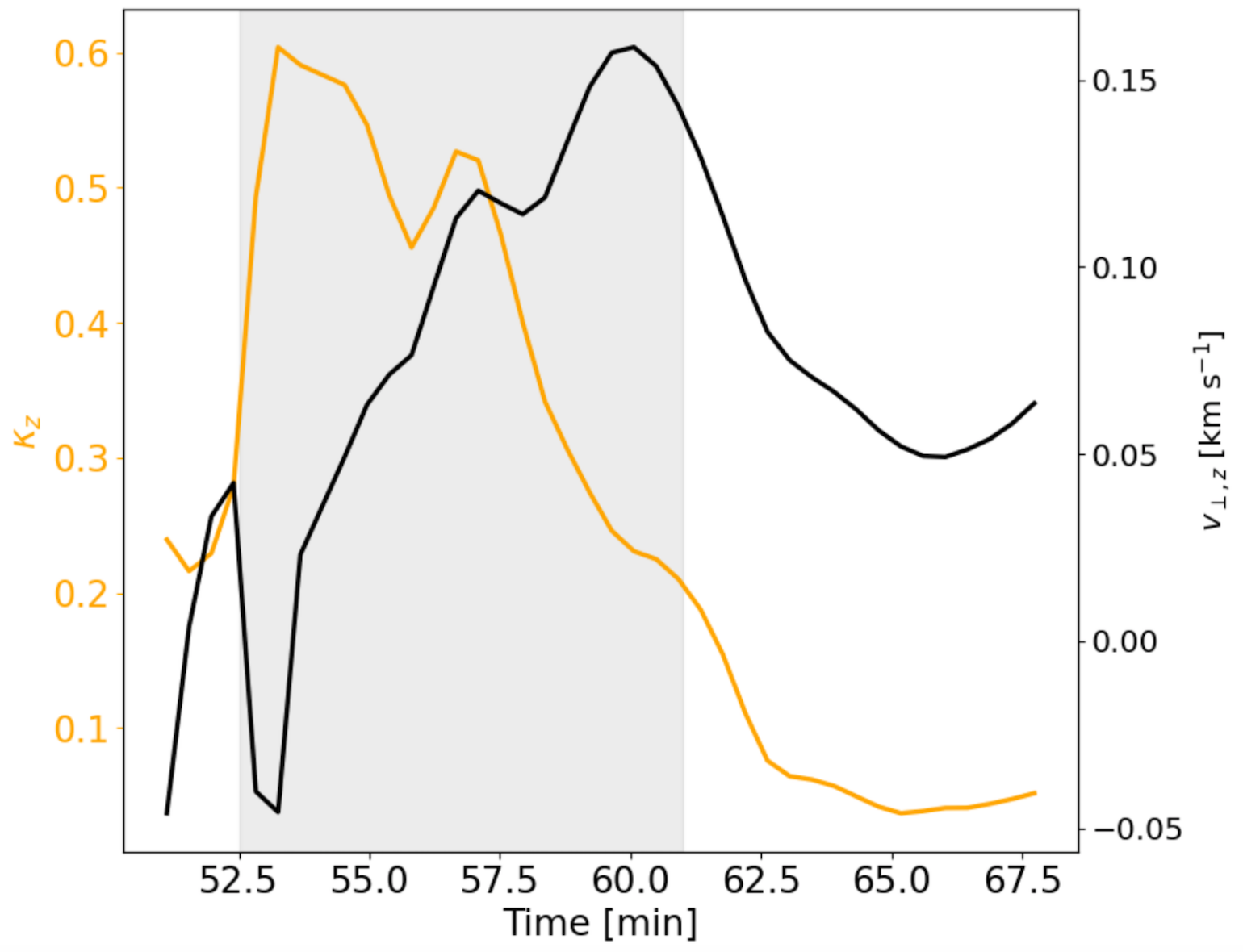}
    \caption{Temporal evolution of the average curvature of the field lines close to the photospheric surface, above the inner bipole PIL (orange curve), and of the vertical  velocity perpendicular to the magnetic field in the same region (black curve). The grey vertical band corresponds to the main cancellation event.}
    \label{fig:vz_kz}
\end{figure}

To understand both the processes that lead to the magnetic flux cancellation discussed in Section \ref{3.2} and its impact on the magnetic field above the flux cancellation region, we analyse the field-line configuration at photospheric heights. At the early stages of the emergence, when the outer polarities first appear, the field lines are nearly straight and have little to no twist or preferred orientation. This absence of twist and rotational shear at this stage allows the magnetic fields to develop an undulatory shape as they rise through the photosphere. When the inner magnetic patches appear, we detect sea-serpent field lines (e.g.\cite{Archontis2010,Pariat2004,Cheung2014}) due to the low twist of the emerging tube. The morphology of this multipolar region is likely driven by the less twisted magnetic field lines, which adopt an undulating shape as a result of the Parker magnetic buoyancy instability. As the emergence progresses, the central polarities begin to converge and interact, while the outer polarities drift outward.

Fig. \ref{lines1} shows the 3D configuration of the emerging magnetic field at four representative snapshots at $\mathrm{t \approx 50,55, 62,65}$ $\mathrm{minutes}$, corresponding to: i) the pre-cancellation phase  (panel a) ii) the onset of flux cancellation (panel b), iii) the end  of the cancellation event (panel c), and finally iv) the post-cancellation event, when the blowout jet has begun (panel d). In Fig. ~\ref{lines1}, only closed field lines with high magnetic twist values are shown, with footpoints that have field strengths exceeding $40~\mathrm{G}$. Following \citet{liu2016}, we calculated the field-line twist number as 

\begin{equation}
T_w = \int \frac{(\nabla \times \mathbf{B}) \cdot \mathbf{B}}{4\pi B^2}\, ds,
\end{equation}

which represents the number of turns a field line makes around its neighbouring field lines. The integral of the
above equation is taken along each traced field line. Field lines with $\mathrm{|T_w| > 1}$ are selected, which ensures that each field line completes at least one full turn. A coherent concentration of such field lines indicates the presence of a significantly twisted magnetic structure. A similar order-unity twist threshold has been widely used in both MHD simulations and non-linear force-free (NLFFF) extrapolation studies. A magnetic flux rope (MFR) is defined as a coherent bundle of field lines winding about a common axis; in this context, field lines with $\mathrm{|T_w| \gtrsim 1}$ complete at least one full turn and therefore define its coherent core. Thus, spatially coherent sets of field lines with $\mathrm{|T_w| > 1}$ are identified here as an MFR. This criterion is consistent with previous MHD models of flux rope formation (e.g. \citet{fan2003,torokkliem,Aulanier2009}) and observational NLFFF studies (e.g. \citet{Guo2010,Inoue2011,liu2016}), where twist values of order unity are used to delineate the flux rope structure. Fig. ~\ref{lines1} includes maps of the squashing factor $\mathrm{Q}$, which  quantifies the degree of distortion of magnetic field line connectivity (\citet{Titov2007}). High values of $\mathrm{Q}$ indicate regions where neighbouring field lines diverge strongly and highlight quasi-separatrix layers that are associated with strong gradients in connectivity and represent sites where magnetic reconnection is likely to occur.

Panel (a) of Fig. \ref{lines1} corresponds to a snapshot prior to the flux cancellation onset and depicts the early appearance of the inner polarities. Above this region concave-upward loops are found, with weak twist values, smaller than unity. 
In between these purple-green lines is a set of yellow-coloured lines, with larger values of twist. Panel (b) corresponds to the early stage of the cancellation, where U loops are still located directly above the cancellation region. The magnetic field lines lean towards the central PIL and still have a concave-up configuration. The colour coding of the twist indicates values close to unity (blue to purple), consistent with the early formation of a weakly twisted structure. The photospheric $\mathrm{Q}$ map reveals a layer of large Q values along the PIL and in between the cancelling magnetic elements suggesting the presence of a preferred site for magnetic reconnection. Once the cancellation begins at $\mathrm{t \approx 53}$ minutes (panel b Fig. \ref{lines1}), the opposite-polarity patches further converge and continue interacting, developing concave-upward structures extending up to $\mathrm{\sim 1.5~\mathrm{Mm}}$. Just before the onset of cancellation, we detect exclusively positive vertical plasma velocities in the layer extending from $\mathrm{z \approx 0}$ up to $\sim 0.5~\mathrm{Mm}$, with maximum values ranging between $\mathrm{0}$ and $1~\mathrm{km\,s^{-1}}$. Around this region small downflows are detected close to the polarity inversion line. For clarity we include Fig. \ref{uloops}, where in the top panel is a close-up of the newly formed U loops (purple lines) located right above the inner bipole PIL. The background shows a map of the parallel current in the middle of the PIL, where we can detect an increase of the current in this region.

{The bottom panel of Fig.~\ref{uloops} shows the top view of the two magnetic polarities at $\mathrm{z \approx 0}$, where the green vectors represent the total horizontal plasma velocity $\mathrm{{u}_h}$, and the solid red line marks the polarity inversion line (PIL). This horizontal velocity field can be physically decomposed into a shearing component parallel to the PIL and a converging component perpendicular to it. Near the PIL, within the regions of intense magnetic flux, the larger arrows reflect enhanced horizontal velocities. Specifically, we observe counter directed motions along the PIL (shearing) as well as plasma flows driving inwards towards the cancellation region (convergence). In particular, focusing on the region surrounding the PIL, the flow vectors over the negative polarity are directed along the $\mathrm{+y}$-direction and slightly to the left ($\mathrm{-x}$), whereas over the positive polarity, they point towards the $\mathrm{-y}$ direction and slightly to the right ($\mathrm{+x}$). This configuration combines anti-parallel flows along the PIL with inwards motions across it, clearly demonstrating how these combined flows force the polarities to come closer while shearing past one another. However, it is important to note that from the direction and orientation of the velocity vectors alone, it is not visually clear which of the two dynamic processes (shearing or converging) is dominant over the other. Overall, these flows work together to both drag the magnetic fields along the PIL and press them against each other.}
\begin{figure}[h]
\hspace{-0.1cm}    \includegraphics[width=9.2cm]{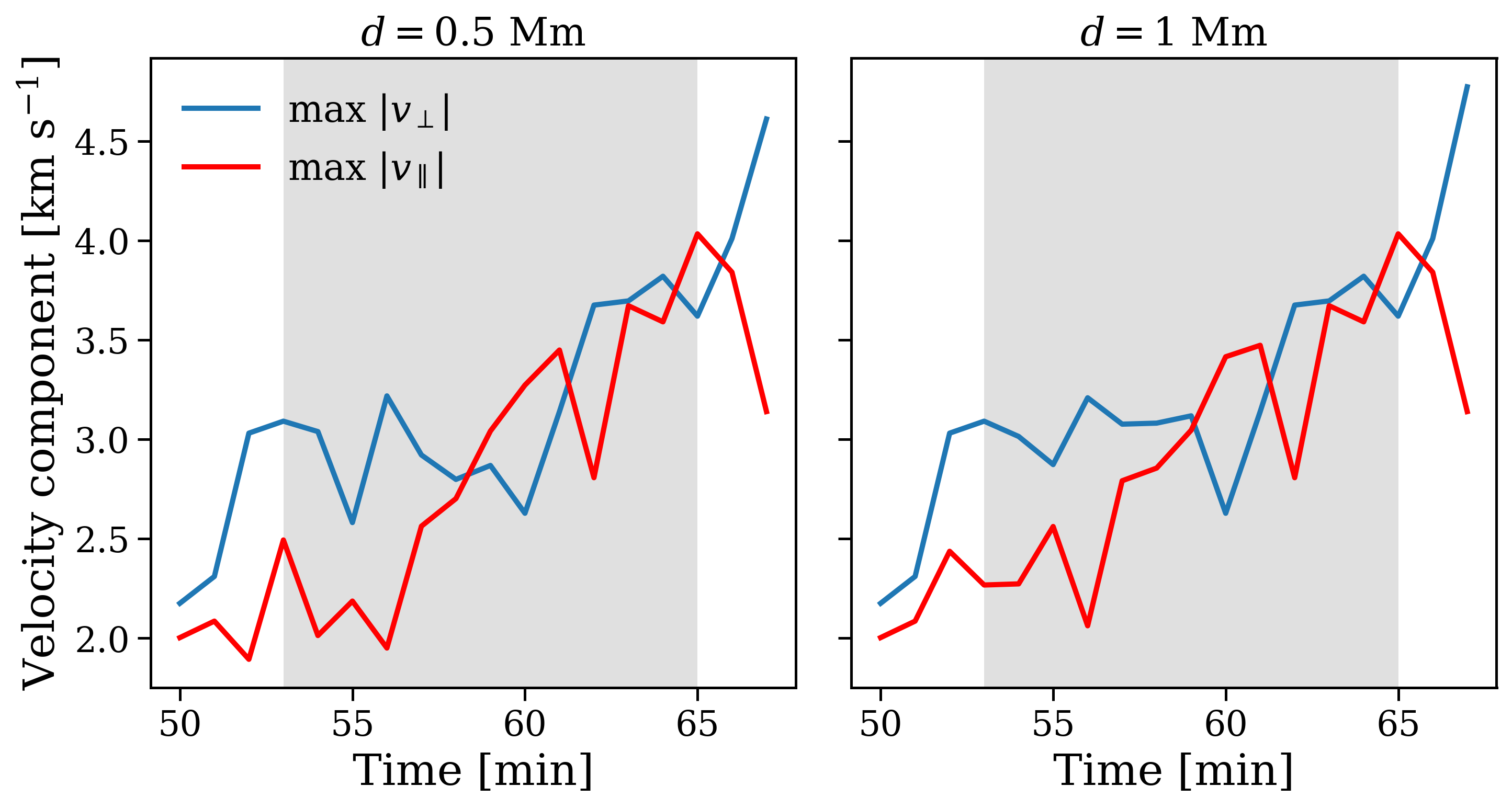}
    \caption{{Temporal evolution of the maximum horizontal velocity components calculated within a distance of $\mathrm{d = 0.5~Mm}$ (left panel) and $\mathrm{d = 1~Mm}$ (right panel) from the main PIL. The blue curves represent the maximum perpendicular (converging) velocity component, $\mathrm{\max |v_\perp|}$, while the red curves denote the maximum parallel (shearing) velocity component, $\mathrm{\max |v_\parallel|}$. The grey shaded region highlights the cancellation phase where the two velocity components compete, showing that neither clearly dominates the other throughout the entire event.}}
    \label{fig:vperpar}
\end{figure}
 \begin{figure}[h]
\centering
  \includegraphics[width=7cm]{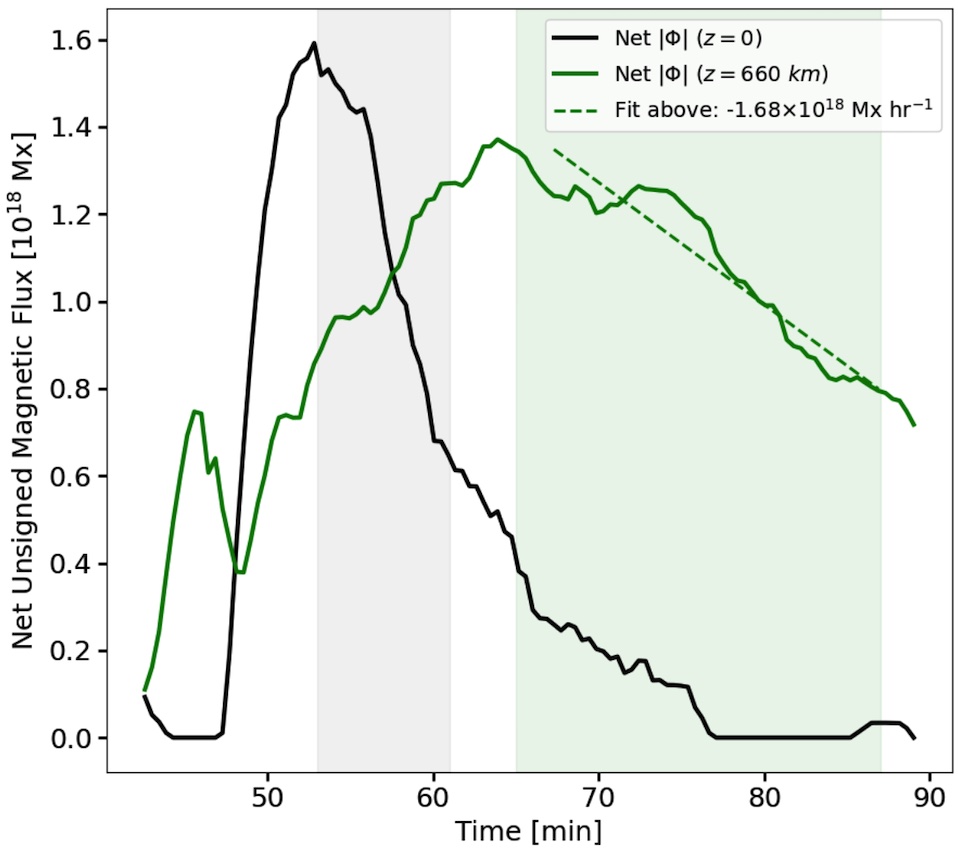}
    \caption{Temporal evolution of the net unsigned magnetic flux at the
photosphere (black curve) and at \(z=660\) km above the photosphere
(green curve). The grey and green shaded regions mark the corresponding
cancellation intervals at the two heights.}
    \label{fig:flux_ab}
\end{figure}
 In order to study the vertical and horizontal motions of the U loops in the cancellation site, we include Fig.  \ref{fig:vz_kz}. Here, we plot the curvature of the field lines (orange), as well as the plasma velocity perpendicular to the magnetic field lines (black). The grey band corresponds to the duration of the cancellation event. We detect positive curvatures as well as slow positive plasma velocities ($\mathrm{\approx 0.15~km~s^{-1}}$) during the cancellation event. {Figure~\ref{fig:vperpar} shows the temporal evolution of the maximum horizontal velocity components, calculated within distances of $d = 0.5~\mathrm{Mm}$ (left panel) and $d = 1.0~\mathrm{Mm}$ (right panel) from the main PIL. At both distances, both the perpendicular (converging) component, $\max |v_\perp|$, and the parallel (shearing) component, $\max |v_\parallel|$, exhibit a generally increasing trend over time. Specifically, around $t = 50~\mathrm{min}$, the velocity components start near $2.0~\mathrm{km~s^{-1}}$ and progressively rise as the system evolves. Within the grey shaded interval ($\mathrm{t = 52}$--$65~\mathrm{min}$), which corresponds to the main dynamic phase of the event, the two components actively compete, with their maximum values fluctuating between approximately $2.5~\mathrm{km~s^{-1}}$ and $4.0~\mathrm{km~s^{-1}}$. Notably, neither component strictly dominates the other throughout this period. Following the end of this phase ($\mathrm{t > 65~\mathrm{min}}$), the converging velocity component $\mathrm{\max |v_\perp|}$ shows a sharp acceleration, peaking at over $\mathrm{4.5~\mathrm{km~s^{-1}}}$. The strong qualitative and quantitative agreement between the two panels demonstrates that this velocity behaviour is highly robust and independent of the chosen distance $\mathrm{d}$ from the PIL.} The larger horizontal velocities indicate that motions in the cancellation region are mainly horizontal rather than vertical. 

 {Figure  \ref{fig:flux_ab} shows the net unsigned flux at photospheric level, as well as $\mathrm{660}$ $\mathrm{km}$ above it. For the photospheric flux (black line) the evolution is the same as in Fig. \ref{fluxcan}. At $\mathrm{z \approx 660~km}$ (green line), we find a delayed increase in flux of $\mathrm{\approx 15}$ $\mathrm{minutes}$. Cancelled flux is also found at this height, as indicated by the green vertical band. The cancellation rate at $\mathrm{660~km}$ is $\approx$ $\mathrm{1.68 \times 10^{18}}$ $\mathrm{Mx~hr^{-1}}$, which is much smaller than the photospheric cancellation rate. Overall, signatures of cancellation are present up to $\mathrm{850}$--$\mathrm{900}~\mathrm{km}$ above the surface.  This difference in cancellation rates could be partially attributed to the fact that magnetic fields become weaker as we move above the photospheric plane. We repeated this calculation at different depths below the surface, and found no clear sign of flux cancellation. We detect a sharp increase and decrease of flux, due to the passage of magnetic fields from sub-photospheric to upper heights, which occurs $\mathrm{\approx2~min.}$ before photospheric cancellation begins.}

\begin{figure}[h]
\hspace{0.7cm}    \includegraphics[width=7cm]{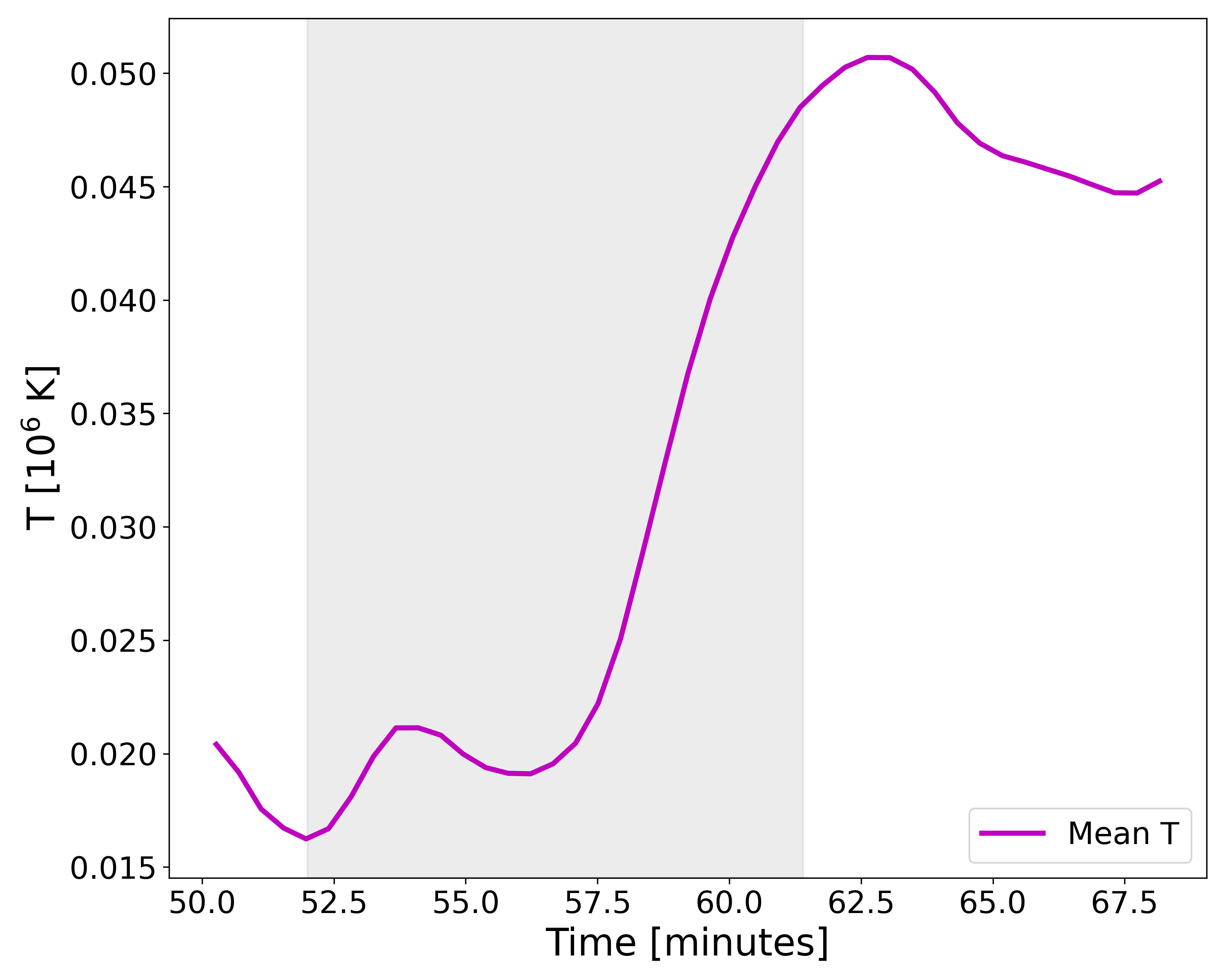}
\caption{Temporal evolution of the plasma temperature above the cancellation site. The purple curve (left axis) shows the average temperature in the low atmosphere ($\mathrm{z \approx 0}$--$2~\mathrm{Mm}$). The shaded region indicates the main flux cancellation phase.}
    \label{fig:Tmean}
\end{figure}

These changes in the magnetic topology during the cancellation event are accompanied by changes in the thermodynamic properties above the cancellation region.  Fig. \ref{fig:Tmean} shows the temporal evolution of the average temperature  (purple line) at photospheric-chromospheric heights ($\mathrm{z \leq 2}~\mathrm{Mm}$). The mean temperature remains relatively low during the early cancellation phase, with values of the order of $\mathrm{2\times10^{4}}$ $\mathrm{K}$, before increasing rapidly after $\mathrm{t\approx59}$ $\mathrm{minutes}$, to approximately  $\mathrm{5\times10^{4}~K}$, coinciding with the onset of the jet phase. This behaviour suggests that the low-atmospheric plasma is heated during the later stages of the cancellation event.

\subsection{Association with flux rope formation}
\label{4.2}
Panel (c) of Fig. ~\ref{lines1} depicts a newly formed twisted rope in the region where cancellation is happening. Shearing motions between the inner magnetic patches lead to magnetic field reconfiguration at the centre of the U loops, resulting in the gradual formation of a flux rope beginning at $\mathrm{t \approx 57}$ minutes. The flux rope continues to expand, with its axis aligned along the $\mathrm{\pm y}$ direction. In this panel enhanced squashing factor values are found, indicating strong gradients in magnetic connectivity around the rope's boundaries. We also observed a substantial increase in the twist of the field lines with respect to the previous panels, confirming the formation of a coherent, highly twisted magnetic structure.

{Panel (d) of Fig. ~\ref{lines1} shows the same configuration immediately after the onset of the blowout jet, when the cancellation process has stopped. The core of the flux rope appears to remain, although the field lines exhibit a weaker average twist compared to the previous panel. This suggests that part of the twist has been redistributed or released during the eruption. Rather than a full large-scale ejective eruption, the MFR undergoes a partial eruption: its outer, more weakly twisted layers are progressively removed through reconnection with the pre-existing ambient field, while its inner core remains more intact. This behaviour is consistent with numerical studies showing that magnetic reconnection between emerging and pre-existing fields can lead to a restructuring of the magnetic topology and repeated energy release, resulting in the formation of successive twisted configurations \citep{arch2010}. Similar reconnection driven interactions between emerging structures and the pre-existing ambient field have also been shown to control the eruptive behaviour of flux ropes in other 3D MHD simulations \citep{archtorok2008,moreno2013}.}

\begin{figure}
    \includegraphics[width=9cm]{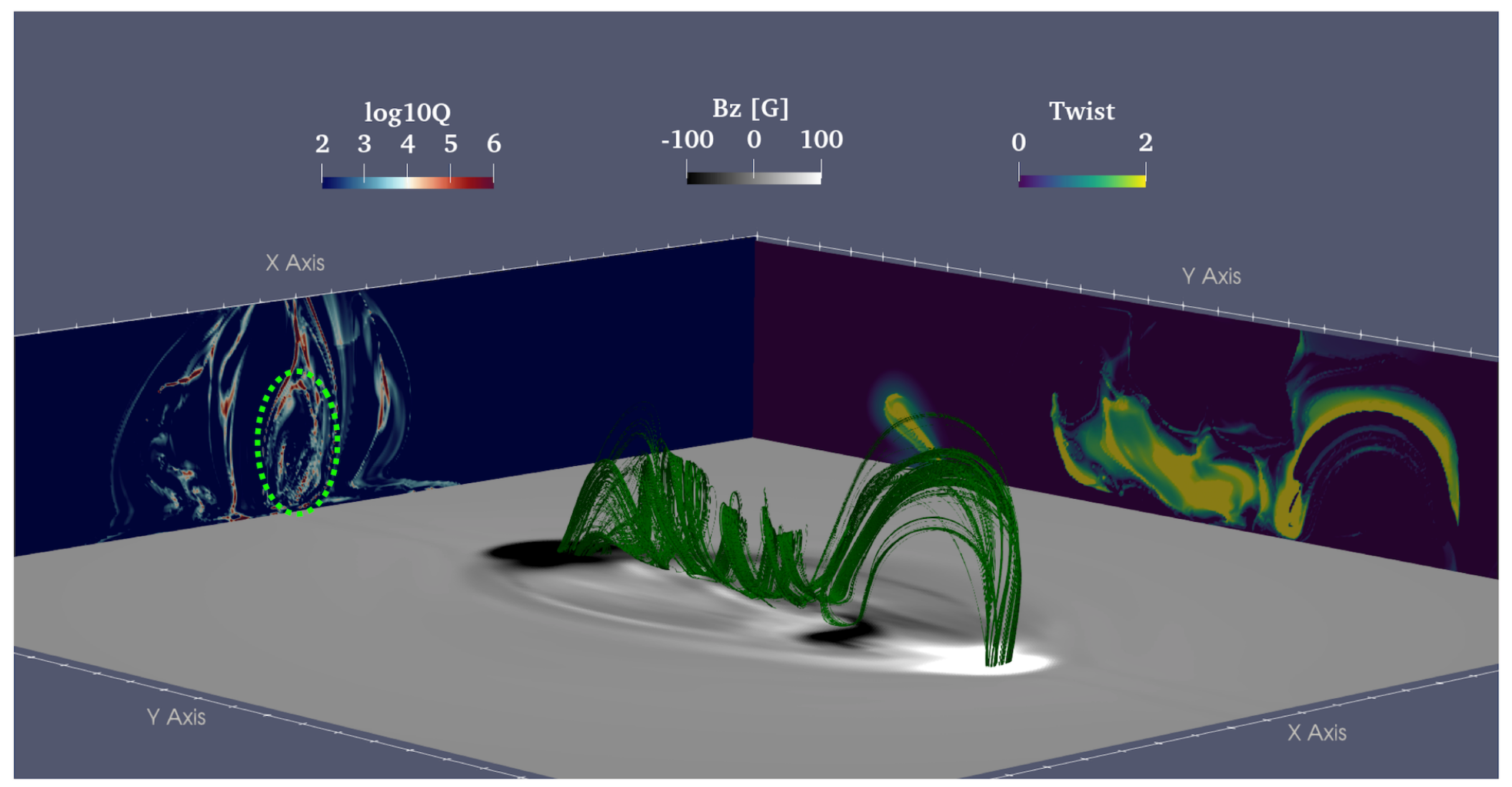}
\caption{Three-dimensional view of the magnetic flux rope during the
middle of the main cancellation phase. The green field lines show the
twisted closed magnetic structure above the photospheric $\mathrm{B_z}$
magnetogram. The side planes show the corresponding $\mathrm{logQ}$ and
$\mathrm{T_w}$ maps. The dotted oval approximately outlines the flux rope
cross-section.} 
   \label{cur}
\end{figure}

To further quantify the properties of the flux rope, its core is isolated using squashing factor maps. Figure~\ref{cur} shows the closed field lines with $\mathrm{T_w>1}$ at $\mathrm{t\approx62}$ minutes, together with a side view of the rope and a twist map taken parallel to its axis. This cut reveals a highly twisted core, shown in yellow colour. A $\mathrm{Q}$ factor map is also shown in a plane perpendicular to the rope axis, allowing the boundaries of the rope cross-section to be identified. The green dotted oval approximately outlines the MFR cross-section. From this analysis, we identify the spine of the rope and estimate its characteristic parameters. The mean twist number inside the rope is $\mathrm{\langle |T_w| \rangle \approx 1.5-1.8}$ over the interval $\mathrm{t = 56-62}$ minutes, indicating a strongly twisted and coherent magnetic structure.

{
Subsequently, we calculated the axial flux in the region where the MFR forms. Since the rope forms above the PIL, we followed the evolution of the PIL and use it as a proxy of  the flux rope axis direction. We then took a cross-section perpendicular to the axis of the rope, with a radius of $\approx\mathrm{1.5~Mm}$, and calculate the magnetic flux through it. At the beginning of the cancellation phase, the rope is not well aligned with the \(y\)-axis. For this reason, we used the magnetic field along the direction of the rope, which is the projection of $\mathrm{B_y}$. Initially the inclination is $\mathrm{27^\circ}$ at $\mathrm{t\approx 52}~\mathrm{min}$. Later, the rope becomes almost aligned with the \(y\)-axis, with the inclination decreasing to about $\mathrm{6^\circ}$ at $\mathrm{t\approx 62}~\mathrm{min}$. During this time interval, the axial flux increases progressively, as  more magnetic flux is transferred into the rope. We now define the fraction $\mathrm{f = \frac{\Delta \Phi_{\mathrm{axial}}}{\Phi_{\mathrm{canc}}}},
$ where $\Delta \Phi_{\mathrm{axial}}$ is the variation of the axial magnetic flux, and $\Phi_{\mathrm{canc}}$ is the total net-unsigned cancelled flux during the mentioned-above interval, estimated to be $\Phi_{\mathrm{canc}} = 8.2\times10^{17}$ $\mathrm{Mx}$. By comparing the flux variations over the interval $\mathrm{t=52.4}$ to $\mathrm{t=62.2}$ minutes, we found that approximately $\mathrm{22-25 \%}$ of the cancelled flux contributes to the increase of the axial flux of the rope. This behaviour confirms that flux cancellation plays a role in building up the flux rope.}
\subsection{Association with the blowout jet}
\label{4.3}

To study the connection between photospheric flux cancellation and the observed jet activity, we examined the temporal evolution of the magnetic flux components in the system. As mentioned earlier, the standard jet ($\mathrm{t = 51 min.}$) begins
slightly before the main cancellation phase and is followed by a
more energetic blowout jet ($\mathrm{t = 64 min.}$) Since the standard jet
begins before the cancellation phase, it shows no direct association with it. To quantify the impact of flux cancellation on the coronal magnetic structure, we computed the evolution of the open and closed twisted magnetic flux. We focused on field lines with $\mathrm{|T_w|>1}$ and footpoints with $\mathrm{|B_z|>40}$ $\mathrm{G}$, distinguishing between open and closed magnetic connectivity within the full numerical domain.

\begin{figure}
\hspace{0.5cm}
    \includegraphics[width=7cm]{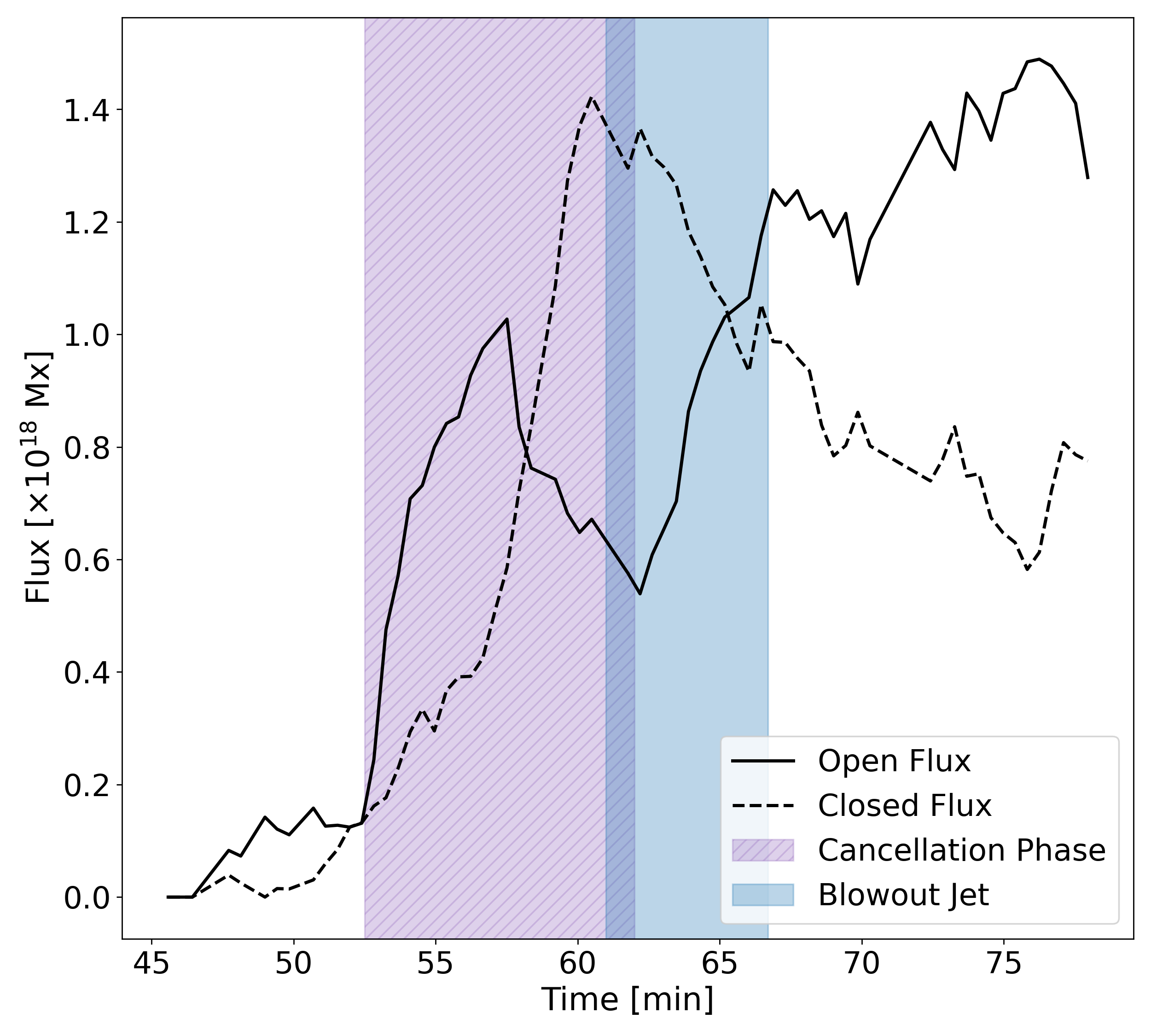} 
    \caption{Temporal evolution of open twisted flux and closed twisted flux (solid and dashed black lines, respectively) calculated for $\mathrm{|T_w|>1}$ and for the whole atmosphere. See text for explanations. The purple vertical band marks the interval of the main cancellation event, while the blue band indicates the time period of the blowout jet.}
 \label{o-c-flux}
\end{figure}

Figure~\ref{o-c-flux} shows the temporal evolution of the open 
(solid curve) and closed (dashed curve) twisted flux. The purple shaded 
region marks the interval of the main cancellation event, while the blue 
shaded region corresponds to the blowout jet. During the cancellation 
phase, the closed twisted flux increases steadily, indicating the gradual 
build-up of the flux rope as magnetic flux is transferred into a coherent, 
highly twisted closed structure. Between $\mathrm{t=55}$ and 
$\mathrm{t=65}$ min, the closed twisted flux changes from 
$\mathrm{9.1\times10^{17}}$ $\mathrm{Mx}$ to 
$\mathrm{1.3\times10^{18}}$ $\mathrm{Mx}$, supporting the interpretation 
that the flux rope is gradually formed during this phase. At the same time, the open twisted flux also increases, suggesting that 
part of the magnetic field reconnects with the surrounding ambient field 
during the standard jet phase. Near the end of the cancellation phase, at 
$\mathrm{t\approx61}$--$62$ min, the closed twisted flux reaches its 
maximum value, while the open twisted flux decreases. This reflects the 
formation of a predominantly closed magnetic configuration. At the onset of the blowout jet, at $\mathrm{t\approx63}$ min, the 
behaviour changes abruptly: the closed twisted flux shows a steep decrease, 
while the open twisted flux increases significantly. This indicates that 
part of the previously closed twisted magnetic field reconnects and opens 
further. During this phase, the closed field-line topology is disrupted, 
as the twisted flux rope discussed in the previous section erupts and 
drives the blowout jet.

\section{Conclusions and discussion}
\label{5}
In this work, we performed a three-dimensional resistive MHD simulation of the emergence of a horizontal magnetic flux tube with a non-uniform twist profile in a magnetised solar atmosphere. The aim of the study was to investigate how magnetic flux cancellation naturally develops during flux emergence and to examine its role in the formation of twisted magnetic structures and jet activity. Our main results are summarised as follows:

\begin{enumerate}
\item We detected a flux cancellation event between two small polarities in the centre of our emerging system with a duration of almost $\mathrm{12}$ $\mathrm{minutes}$, and we found a cancellation rate of $\mathrm{1.1 \times 10^{19}}$ $\mathrm{Mx~hr^{-1}}$.  
\item Approximately 10\% of the total flux of the inner emerging bipole was involved in the cancellation event.
\item We attribute the flux cancellation event to the magnetic reconnection between converging opposite polarities at the PIL driven by strong horizontal flows. 
\item Above the cancellation site we found the formation of a twisted flux rope midway through the cancellation process. This rope eventually erupted during a blowout jet that appeared at the end of the cancellation event. We quantified the role of the cancelled flux in the development of the twisted rope, and we found that $22-25$\% of the cancelled flux contributes to the axial flux of the rope.
\item The standard jet does not appear to be directly driven by the main
cancellation event.
\end{enumerate}

A key characteristic of the emerging region in our simulation is the development of a complex multipolar photospheric structure. Instead of forming a simple bipolar region, the magnetic field produces several secondary opposite-polarity elements between the main polarities. This behaviour is directly related to the non-uniform twist profile of the initial flux tube. Mildly twisted fields tend to develop undulations as they approach the photosphere, producing alternating polarity patches interpreted as the surface manifestation of serpentine magnetic field lines. Similar multipolar configurations have been reported in previous numerical studies of weakly twisted flux emergence (e.g. \cite{Archontis2010, Toriumi2011, Syntelis2019}). In our simulation these secondary polarities play a crucial role in driving flux cancellation. The emergence of new flux in pre-existing active regions is a very common scenario in solar observations \citep{zuccarello2008,lites2010,zhang2012,guglielmino2018,murabito2025}. We found additional smaller magnetic poles between the main polarities of the pre-existing active region. These small-scale opposite-polarity pairs are known as moving dipolar features (MDFs), a term introduced by \citet{bernasconi2002} to describe transient dipoles that appear within developing active regions. 
\citet{centeno2012} used HMI/SDO data to study the naked emergence of active regions, meaning the isolated flux appearing in regions with relatively quiet magnetic surroundings. They studied the emergence of a dipole inside an active region, which can be seen in Figure 10 of their work. In the central emergence zone, between the two main polarities, two small MDFs appeared, approached each other, and eventually cancelled. This behaviour contrasts with the findings of \citet{bernasconi2002}, where MDFs drifted outwards towards the stronger pre-existing polarities and disappeared through cancellation with the surrounding field. In the \citet{centeno2012} case, the interaction remained confined to the emergence site, and redshifted Doppler signatures were detected at the MDF locations, indicating plasma draining along low-lying field lines, consistent with shallow U loops that possibly undergo partial submergence.

In our simulation, the cancellation event occurs when the inner opposite-polarity magnetic patches converge along an internal PIL. By analysing the temporal evolution of the magnetic flux within the interacting region, we estimated a total cancelled magnetic flux of approximately $\mathrm{\Phi_{\mathrm{canc}} \sim 10^{18}}$ $\mathrm{Mx}$ and a characteristic cancellation rate of $\mathrm{\sim 10^{19}}$ $\mathrm{Mx~ hr^{-1}}$. The close balance between the positive and negative flux decrease confirms that this process corresponds to genuine flux cancellation rather than simple advection of magnetic flux. Our estimation of the cancellation rate is in agreement with previous observational and numerical studies, which typically report values in the range of $\sim 10^{18}$--$10^{19}\ \mathrm{Mx\ hr^{-1}}$ (e.g.  \cite{ park2009,panesar2018,Kaithakkal2019,ledvina2022,patsourakos2025,moreno2025}).

We also performed a brief analysis of the secondary flux cancellation events, which we present in Appendix A. 
In contrast to the main cancellation episode, these events involve largely unequal amounts of positive and negative magnetic flux, indicating partial cancellation in the sense of \citet{2007ApJ...666..576D}. They are also characterised by lower 
cancellation rates, of the order of $\sim2.6\times10^{18}$ to 
$\sim5\times10^{18}~\mathrm{Mx\,hr^{-1}}$, and by shorter durations 
($\sim6$--$8$ minutes), compared to the more extended and intense main event. For these secondary events, we find no association with any of the major dynamical responses of the system. 

The magnetic topology and velocity patterns provide further insight into the physical mechanism responsible for the disappearance of flux. During the cancellation phase, we detected the formation of concave-up magnetic loops above the internal PIL. Due to large horizontal flows, reconnection is likely to occur at the centre of the loops. Additionally, an important result of our analysis is that this cancellation process contributes to the formation of a coherent magnetic flux rope above the internal PIL. As the opposite polarities interact, magnetic flux is transferred into a twisted core structure, which reaches mean twist values of $\mathrm{\langle |Tw| \rangle \approx 1.5}$--$\mathrm{1.8}$, despite the relatively weak initial twist of the emerging flux tube. By comparing the magnetic flux components with the total cancelled flux, we found that the axial magnetic flux increases  during the cancellation phase.

{Although the mean twist of the formed magnetic flux rope reaches relatively high values ($\mathrm{\sim 1.5\text{--}1.8}$ turns), approaching the kink instability threshold of $\mathrm{\Phi/2\pi \approx 1.75}$ reported by \citet{Torok2004}, the MFR does not undergo a kink instability. To understand this behaviour, several differences between our simulation and ideal models must be taken into account. In \citet{Torok2004}, the initial coronal loop is modelled as a pre-existing force-free field with a purely axisymmetric geometry, allowing the twist to be defined via the ratio $\mathrm{\Phi/2\pi}$ \citep{Hood1979}, where $\mathrm{\Phi}$ is the total twist angle. In contrast, our field was generated dynamically via flux emergence, it is non-force-free, and has a non-axisymmetric configuration. Moreover, the twist was computed using a current-based formulation \citep{Berger2006, liu2016}, which precludes a direct quantitative comparison between the two metrics. Furthermore, as emphasised by \citet{Torok2004}, the critical twist threshold for the onset of the kink instability depends strongly on the flux tube aspect ratio, with more elongated structures requiring substantially higher twist levels to become kink unstable. 

The dynamic evolution of the system also revealed two jet-like events in the corona. The first event corresponds to a standard reconnection jet, and it occurred slightly before the onset of the cancellation episode. This jet appears to be driven primarily by magnetic reconnection between the emerging magnetic field and the open coronal field, rather than by the cancellation process itself. In contrast, the second jet occurred after the flux rope had formed and expanded. During this eruptive phase, the closed twisted magnetic flux decreased, while the open twisted flux increased, indicating that part of the previously closed twisted field reconnected with the surrounding coronal magnetic field and became open. This behaviour is consistent with the eruption of a flux rope structure that releases part of its stored magnetic twist into the overlying corona. Overall, our results provide numerical evidence that photospheric flux cancellation in emerging magnetic regions can naturally drive the formation of coherent magnetic flux ropes. 

We also note that flux cancellation is not unique to the case presented here. It has been observed in other non-uniform twist simulations as well. In those cases, similar interactions between opposite polarity magnetic elements are found, suggesting that flux cancellation is a generic feature of the emergence process, which depends on the specific radial twist profile. A comparative analysis of cancellation across different twist configurations is beyond the scope of the present work and will be addressed in a future study. We further remark that the identification of cancellation events in this work 
is based primarily on visual inspection of the photospheric magnetograms, together with the temporal evolution of magnetic flux. Given the complex structure of the photospheric magnetic field, it is possible that weaker or more transient 
events have not been captured. Future work could benefit from the use of 
automated feature tracking techniques such as solar feature tracking (i.e. SoFT; \cite{Berretti}), which would allow for a more direct identification of flux cancellation.

\begin{acknowledgements}
{The authors would like to thank
the referee for useful comments and suggestions.}
The authors acknowledge support by the European Research Council (ERC) Synergy Grant ‘Whole Sun’ (No. 810218). This work was supported by computational resources granted by the National Infrastructures for Research and Technology S.A. (GRNET S.A.) at the National HPC facility ARIS, under the project MFES.
\end{acknowledgements}

\bibliographystyle{aa}  
\bibliography{references}

\begin{appendix}

\section{Secondary flux cancellation events}
\label{3.5}

The simulation reveals frequent 
flux cancellation events in the photosphere, reflecting the complex magnetic configuration that develops during flux emergence. In addition to the main cancellation episode analysed in Section~3, we identify two additional regions where opposite polarity elements converge and cancel. Unlike the primary event, where the interacting elements cancel close to their emergence site, the events examined here involve magnetic patches that have migrated several megametres before interacting. Figure~\ref{2-3} shows the magnetogram (left panel) at $\mathrm{z=0}$ $\mathrm{Mm}$ and the corresponding $\mathrm{|J|/|B|}$ map (right panel) at $\mathrm{t\approx85}$ $\mathrm{minutes}$. The normalised current density highlights regions of strong current concentration, indicating the presence of narrow current layers and magnetic reconnection along the local polarity inversion line (PIL).

Two regions of interest are marked in the figure: the second cancellation event (yellow box) and the third event (pink box). In the following subsections, we analyse each site separately, examining the temporal evolution of the unsigned and signed magnetic flux, together with the co-temporal enhancement of $\mathrm{|J|/|B|}$ as an indicator of localised current layers and reconnection at the PIL.

\begin{figure}[H]
\hspace{-1.2cm}
\includegraphics[width=10.5cm]{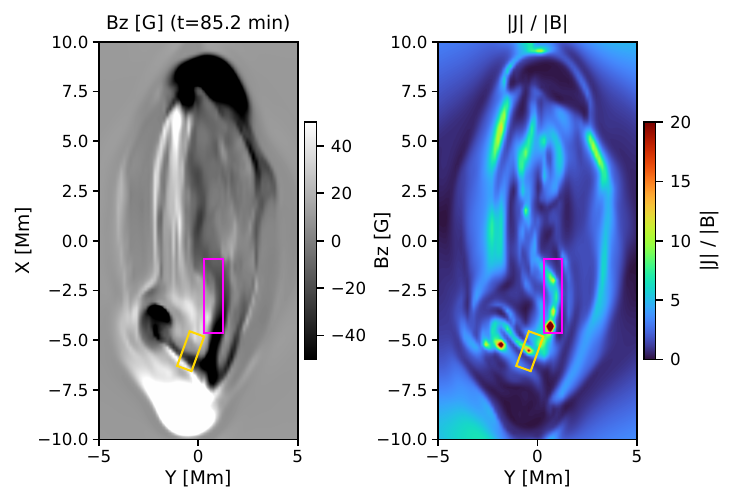}
\caption{Left panel: magnetogram of $\mathrm{B_z}$ at $\mathrm{t\approx85}$ $\mathrm{minutes}$ for $\mathrm{z=0}$ $\mathrm{Mm}$. Right panel: map of $\mathrm{|J|/|B|}$ at the same time. The yellow box encloses the region where the second flux cancellation event occurs, while the pink box encloses the region corresponding to the third cancellation event.}
\label{2-3}
\end{figure}

\subsection{Second event}
\label{3.5.1}

The second cancellation event occurs later in the simulation, at $\mathrm{t\approx78}$ minutes, and lasts for approximately $\mathrm{8}$ minutes. In contrast to the main event, the interacting magnetic elements are not newly emerged flux concentrations. Instead, the small negative polarity patch has migrated by $\mathrm{\sim4}$ $\mathrm{Mm}$ from its emergence location, following the horizontal expansion of the major positive polarity. It eventually encounters a smaller positive patch attached to the outer polarity, which moves outwards more slowly. Thus, both elements are pre-existing magnetic fragments that have already interacted with their surroundings before converging.

Figure~\ref{region2} shows the temporal evolution of the unsigned (black), positive (orange), and negative (purple) magnetic flux within the selected region. Unlike the main cancellation event (Fig. ~\ref{fluxcan}), the positive flux dominates from the beginning, as it pre-exists in the region while the negative element moves into it. Once the opposite polarities converge, the net magnetic flux decreases significantly, with a rate of approximately $\mathrm{5\times10^{18}}\,\mathrm{Mx\,hr^{-1}}$. The magnetogram and $\mathrm{|J|/|B|}$ map at $\mathrm{t\approx85}$ minutes (Fig. ~\ref{2-3}) reveal the interaction: the negative polarity appears embedded within the extended positive field, while the same region coincides with a narrow elongated ridge of enhanced current density. The highlighted area marks the core of this current sheet, indicating that the flux decrease is driven by localised magnetic reconnection at photospheric heights.

\begin{figure}
\includegraphics[width=9.1cm]{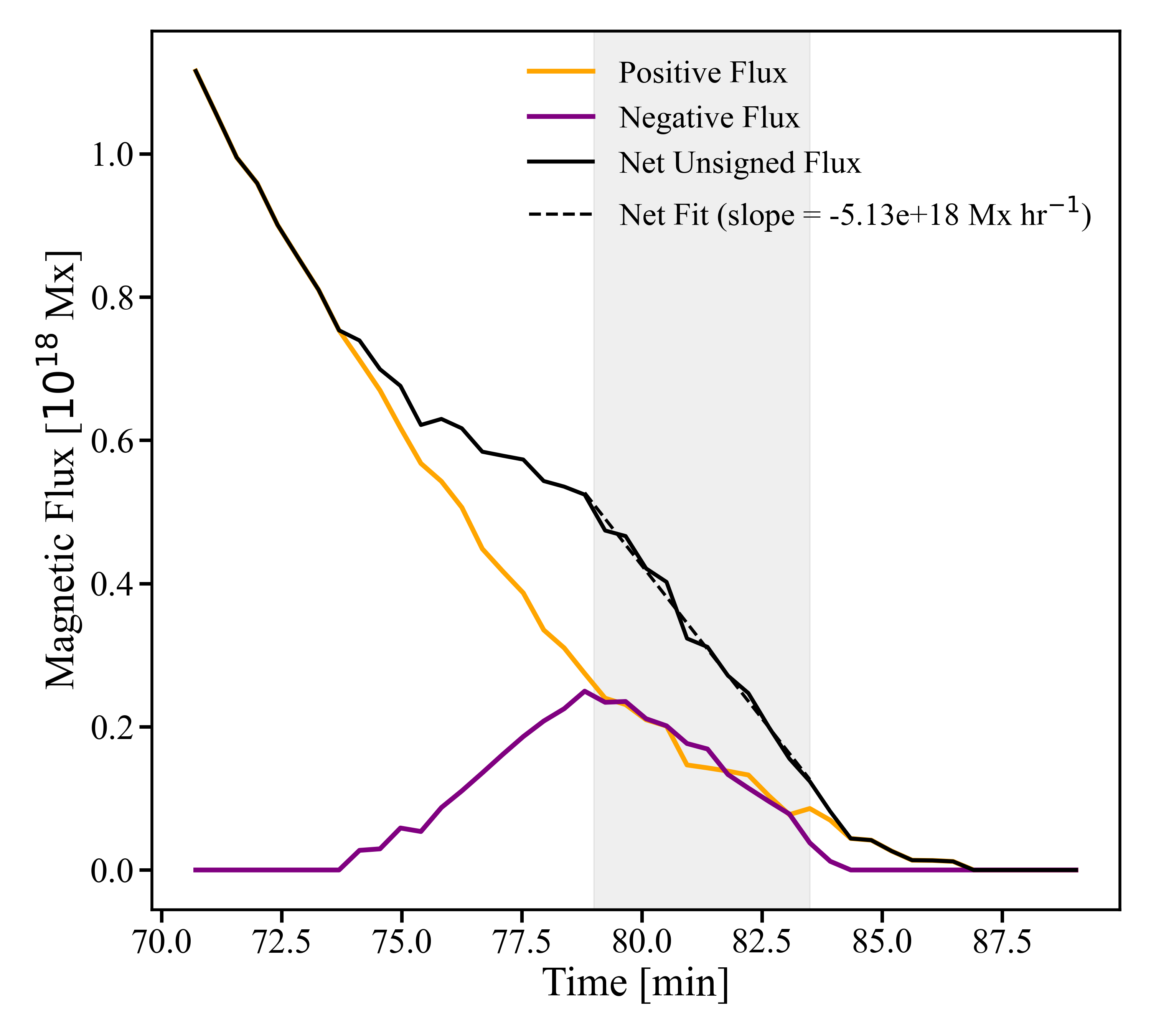}
\caption{Second flux cancellation event: temporal evolution of the net unsigned flux (black curve), the positive flux (orange curve), and the negative flux (purple curve) for the region enclosed by the yellow box in Fig. ~\ref{2-3}. The vertical grey band indicates the time interval during which flux cancellation occurs.}
\label{region2}
\end{figure}

\subsection{Third event}
\label{3.5.2}
The third event occurs within a similar time interval ($\mathrm{t\approx78}$ minutes) but lasts for only about $\mathrm{6}$ minutes. In this case, the interacting magnetic fragments are elongated structures, whose lower parts are advected towards the negative-$\mathrm{\mathrm{y}}$ direction by local horizontal flows, forming narrow stripe-like features of opposite polarity.

Initially, the magnetic fluxes are not balanced, as the negative element occupies a larger area and consists of stronger vertical magnetic fields. This is shown in Fig. ~\ref{region3}, where the net flux initially follows the negative flux (purple--black curve). As the fragments approach each other, the polarity inversion line becomes nearly parallel to their elongation, indicating a combination of converging and shearing motions. In the magnetogram, this interface appears as a sharp boundary between opposite polarities, while the corresponding $\mathrm{|J|/|B|}$ map shows a thin filamentary region of enhanced current density along the same boundary. During the cancellation phase, the net magnetic flux decreases at a rate of approximately $\mathrm{2.6\times10^{18}}\,\mathrm{Mx\,hr^{-1}}$. 

The elongated geometry and shear suggest that reconnection could occur along an extended current sheet rather than at a single localised point. This can explain the short duration ($\mathrm{\sim6}$ minutes) yet efficient flux cancellation. No large-scale jet activity is detected, indicating that this event remains confined to the low atmosphere.
\begin{figure}[H]
\hspace{-0.3cm}
\includegraphics[width=9cm]{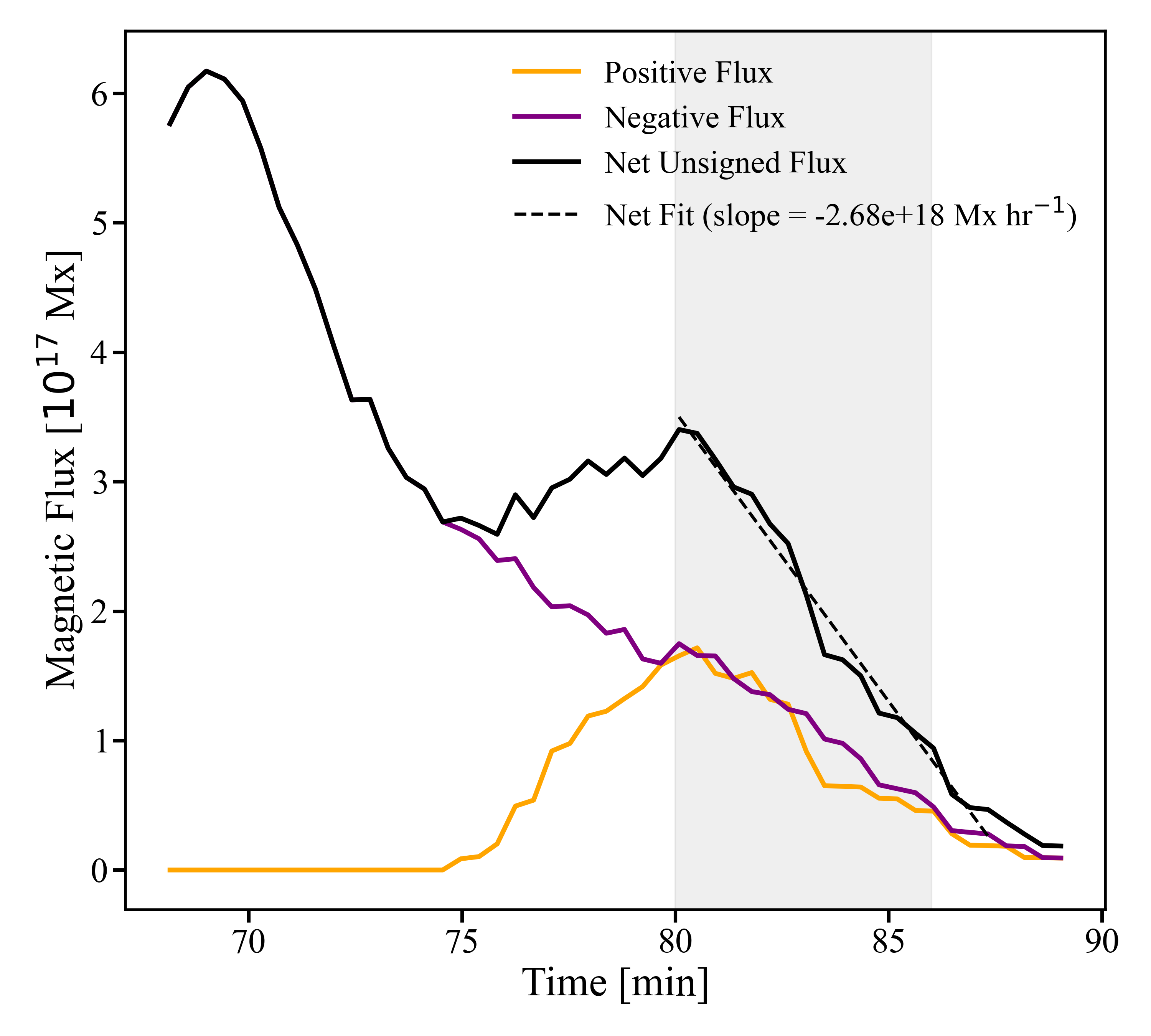}
\caption{Third flux cancellation event: temporal evolution of the net unsigned flux (black curve), the positive flux (orange curve) and the negative flux (purple curve) for the selected region shown inside the pink box in Fig. ~\ref{2-3}. The grey vertical band marks the time interval during which cancellation occurs.}
\label{region3}
\end{figure}

Compared with the main cancellation episode discussed in Section~3, the secondary events analysed here are weaker and shorter lived. While the primary event involves larger magnetic elements interacting close to their emergence site and contributes to the formation of the flux rope, the events presented in this appendix involve smaller fragments that have migrated through the photospheric flows before interacting. The presence of multiple cancellation regions is a natural consequence of the multipolar configuration produced by the emergence of the non-uniformly twisted flux tube. As the magnetic field expands horizontally, fragments are advected by photospheric flows forming numerous small-scale polarity inversion lines where reconnection can occur. Such partial cancellation events are consistent with serpentine magnetic structures formed during flux emergence, when weakly twisted field lines develop undulations near the photosphere. Although these events are smaller than the main cancellation episode, they highlight the dynamic and fragmented nature of the photospheric magnetic field.
\end{appendix}

\end{document}